\documentclass[11pt]{article}

\usepackage{hyperref}

\usepackage{amsmath}
\usepackage{amsthm}
\usepackage{amssymb}
\usepackage{amsfonts}
\usepackage{mathtools}
\usepackage{graphicx}
\usepackage{hyperref}
\usepackage{verbatim}
\usepackage{bbm}
\usepackage{color}
\usepackage{xspace}
\usepackage{multirow}
\usepackage{longtable}
\usepackage{url}
\usepackage{microtype}
\usepackage{subfigure}
\usepackage{booktabs}
\usepackage{comment}
\usepackage{algorithm2e}
\usepackage{natbib}
\usepackage{bm}
\usepackage[margin=1in]{geometry}
\usepackage{float}
\usepackage[capitalize,noabbrev]{cleveref}
\usepackage{authblk}

\theoremstyle{plain}

\theoremstyle{definition}

\theoremstyle{remark}

\usepackage[textsize=tiny]{todonotes}

\def\R{\mathbb{R}}

\newcommand{\One}[1]{{\mathbbm{1}}\left\{{#1}\right\}}
\newcommand{\norm}[1]{\lVert{#1}\rVert}

\newcommand{\EE}[1]{\mathbb{E}\left[{#1}\right]} 

\newcommand{\bmbeta}{\bm{\beta}}

\def\R{\mathbb{R}}
\def\Z{\mathbf{Z}}
\def\W{\mathbf{W}}

\newcommand{\X}{\mathbf{X}}
\newcommand{\Y}{\mathbf{Y}}

\newcommand{\bmepsilon}{\bm{\epsilon}}

\DeclareMathOperator{\diag}{diag}

\newcommand{\indep}{\perp \!\!\! \perp}
\newcommand{\iidsim}{\stackrel{\mathrm{iid}}{\sim}}
\usepackage{color}
\usepackage{graphicx}
\usepackage{amsmath}
\usepackage{hyperref}
\usepackage{verbatim}
\usepackage{adjustbox}
\usepackage{amssymb}
\usepackage{amsfonts}
\usepackage{mathtools}

\usepackage{bbm}
\usepackage{color}
\usepackage{xspace}
\usepackage{graphicx}
\usepackage{multirow}
\usepackage{longtable}
\usepackage{url}
\makeatletter
\g@addto@macro{\UrlBreaks}{\UrlOrds}
\makeatother

\def\R{\mathbb{R}}

\def\R{\mathbb{R}}
\def\Z{\mathbf{Z}}
\def\W{\mathbf{W}}

\renewcommand{\thetable}{S\arabic{table}}

\title{Variable selection with FDR control for noisy data – an application to screening metabolites that are associated with breast and colorectal cancer}

\author[1]{Runqiu Wang\thanks{ruwang@unmc.edu}}
\author[1]{Ran Dai\thanks{ran.dai@unmc.edu}}
\author[2]{Ying Huang}
\author[2]{Marian L. Neuhouser}
\author[2]{Johanna W. Lampe}
\author[3]{Daniel Raftery}
\author[4]{Fred K. Tabung}
\author[1]{Cheng Zheng\thanks{cheng.zheng@unmc.edu}}

\affil[1]{Department of Biostatistics, University of Nebraska Medical Center, Omaha, Nebraska, U.S.A.}

\affil[2]{Division of Public Health Sciences, Fred Hutchinson Cancer Center, Seattle, Washington, U.S.A.}

\affil[3]{Department of Anesthesiology and Pain Medicine, University of Washington, Seattle, Washington, U.S.A.}

\affil[4]{Department of Internal Medicine, College of Medicine and Comprehensive Cancer Center, The Ohio State University, Columbus, Ohio, U.S.A.}

\begin{document}

\maketitle

\begin{abstract}

The rapidly expanding field of metabolomics presents an invaluable resource for understanding the associations between metabolites and various diseases. However, the high dimensionality, presence of missing values, and measurement errors associated with metabolomics data can present challenges in developing reliable and reproducible methodologies for disease association studies. Therefore, there is a compelling need to develop robust statistical methods that can navigate these complexities to achieve reliable and reproducible disease association studies. In this paper, we focus on developing such a methodology with an emphasis on controlling the False Discovery Rate during the screening of mutual metabolomic signals for multiple disease outcomes. We illustrate the versatility and performance of this procedure in a variety of scenarios, dealing with missing data and measurement errors. As a specific application of this novel methodology, we target two of the most prevalent cancers among US women: breast cancer and colorectal cancer. By applying our method to the Women’s Health Initiative data, we successfully identify metabolites that are associated with either or both of these cancers, demonstrating the practical utility and potential of our method in identifying consistent risk factors and understanding shared mechanisms between diseases.\\

\end{abstract}

\section{Introduction}
\label{sec:intro}
Breast cancer (BC) and colorectal cancer (CRC) have a high incidence rate, ranking as the highest and third highest among women in the US, respectively \citep{american_cancer_society_2020}. Both cancers share several diet and lifestyle risk factors \citep{wcrf_aicr_2018}. According to the World Cancer Research Fund (WCRF)/American Institute for Cancer Research (AICR) Expert Panel, there is ``convincing" evidence that adult weight gain and excess body fat increase the risk for post-menopausal BC and CRC, and that physical activity reduces the risk for both cancers. Furthermore, alcoholic drinks have been found to increase the risk of post-menopausal BC and CRC. Higher intakes of red meat, animal fats, and refined carbohydrates have been associated with increased risks of both BC and CRC, whereas fruits, vegetables, whole grains, and dietary fiber tend to be linked with reduced risk \citep{yusof2012, xiao2019, putri2013}. However, the WCRF/AICR Expert Panel's classification for the level of evidence supporting the associations for these dietary components remains ``suggestive" or ``probable" rather than ``convincing", except for increased intakes of processed meat and the risk of CRC \citep{wcrf_aicr_2018}. Given this context, there is a crucial need to identify ``convincing" evidence for risk factors associated with BC and CRC. Additionally, it is essential to study the common risk factors for these two prevalent cancers to better understand their shared underlying mechanisms and develop effective prevention strategies.

Metabolomics, the extensive analysis of small molecules in organisms \citep{nannini2020}, reflects both internal cellular processes and external exposures, making it a sensitive tool for tracing pathways associated with chronic diseases like cancer. Despite its use in early cancer detection \citep{cheung2019,yang2020,his2019,zhu2014}, few studies have systematically explored metabolomics in relation to BC and CRC. Identifying metabolomic components could uncover new BC pathways. For example, studies using the European Prospective Investigation into Cancer (EPIC) and the Prospective Lung, Colorectal and Ovarian Cancer (PLCO) cohorts found certain plasma components and pre-diagnostic diet-related metabolites significantly associated with BC risk \citep{his2019, playdon2017}. Further uncovering of BC and CRC-related features could illuminate disease-related biological pathways, enhancing prevention and treatment strategies.


Developing screening methods for metabolomics data presents several challenges due to their inherent characteristics. Metabolomics data are high-dimensional, often containing missing values and measurement errors. The high dimensionality of metabolomics data is a double-edged sword: while it encompasses all potential components associated with the disease, the majority of these components are unrelated, introducing a significant amount of noise. Missingness and measurement errors are inevitable when dealing with large-scale data collection. These factors pose considerable challenges in developing screening methods for metabolomics data that offer reproducibility guarantees. To address these issues, robust and innovative approaches must be designed, which can account for the complexities and limitations associated with high-dimensional, noisy data while still providing accurate and reliable results.


Measurement errors and missing data frequently arise in complex data analysis tasks, posing challenges that must be carefully addressed when designing variable selection procedures. Naive approaches often lead to problematic results. {\color{black}For example, using complete case analysis removing all samples with any missing data leads to spurious results in variable selection. Measurement errors lead to inflated estimation error for coefficients and inconsistency in Lasso variable selection procedure \citep{sorensen2015}.} There has been a surge in the literature on variable selection in the presence of missing data and measurement errors. For missing data mechanisms like missing at random (MAR) and missing completely at random (MCAR), researchers have developed imputation-based methods and other techniques \citep{little2002,tsiatis2006,rassler2013} tailored for variable selection purposes \citep{wolfson2011, johnson2008, garcia2009}. In the context of variable selection with measurement errors, several methods have been proposed, including CocoLasso\citep{datta2017}, corrected Lasso \citep{loh2012, sorensen2015}, generalized matrix uncertainty selector \citep{rosenbaum2013}, generalized matrix uncertainty Lasso \citep{sorensen2015}, and generalized Dantzig selector \citep{antoniadis2010}. These advancements demonstrate the ongoing efforts to tackle the challenges posed by missing data and measurement errors in variable selection.


One critical challenge in variable selection is ensuring replicability guarantees. Over the past few decades, a novel measure for Type I error, the False Discovery Rate (FDR), or the expectation of the false discovery proportion (FDP), has been proposed to address this issue. The renowned Benjamini-Hochberg (BH) procedure \citep{benjamini1995} has sparked a new era in multiple hypothesis testing, leading to the rapid development of methods that control FDR. Among these techniques, the Knockoff-based methods \citep{barber2015,barber2019,candes2018} offer several advantages, making it particularly attractive for various applications. Some of the key benefits of the Knockoff methods include: mild assumptions about data structure, allowing for more flexibility in handling diverse datasets; compatibility with a wide array of models and variable selection procedures {\color{black} for both low and high dimensional data}, 
powerful method with finite sample FDR control guarantee, providing reliable results even with limited sample sizes.

 These advantages make the Knockoff methods especially suitable for applications in metabolomic data, which typically exhibit complex correlation structures, high-dimensional features, and unknown signal strength. Furthermore, the Knockoff methods excel at handling arbitrary correlation structures and do not require prior knowledge of signal amplitudes or noise levels, making them a powerful and versatile tools in the realm of variable selection.


\subsection{Prior work}

\paragraph{Knockoff-based methods}
Advancements in multiple testing problems within a single experiment have resulted in the development of powerful knockoff-based methods that provide exact FDR control for selecting features with conditional associations with the response \citep{barber2015, candes2018}. The knockoff filter by \citet{barber2015} offers exact FDR control for linear models without needing detailed model information and has been developed for high-dimensional cases \citep{barber2019}. The Model-X knockoff by \citet{candes2018} extends this to nonlinear models with unknown response distributions, but requires knowledge of $\X$'s distribution. \citet{barber2020} demonstrated that the Model-X knockoff method is robust to errors in estimating the distribution of $\X$, while \citet{huang} relaxed its assumptions, allowing FDR control as long as parametric form of the distribution of $\X$ is known. A number of publications have explored the construction of knockoffs with approximated distributions of $\X$. For instance, \citet{romano2019} developed a Deep knockoff machine using deep generative models, \citet{liu2019} created a Model-X generating method employing deep latent variable models, and more recently, \citet{bates2020} proposed an efficient general metropolized knockoff sampler. \citet{spector2020} suggested constructing knockoffs by minimizing the reconstructability of features. Knockoff-based methods have also been extended to test the intersection of null hypotheses, leading to the development of the group and multitask knockoff methods \citep{dai2016} and prototype group knockoff methods \citep{chen2020}. 

\paragraph{Current advance in FDR control for identifying simultaneous signals}

Simultaneous signal detection has been explored using BH procedure-based methods \citep{heller2014, bogomolov2013, bogomolov2018}, local FDR \citep{chi2008, heller2014b}, and nonparametric approaches \citep{zhao2020}. {\color{black} These methods rely on the independence, or positive regression dependency property of the features, which do not hold for most metabolomics studies.} 
Recently, \citet{li2021} and \citet{dai2021multiple} introduced the multi-enviornment knockoff and the simultaneous knockoff methods for feature selection and identifying consistent associations, potentially useful for detecting mutual BC and CRC risk factors. {\color{black}However, all these methods do not allow the existence of missing data or measurement errors, which presents an important and unavoidable issue in metabolomic data analyses.}

\subsection{Our contributions}
In this paper, we propose a knockoff-based procedure to establish FDR control in selecting (mutual) signals when there are missing data and error-prone data assuming very general conditional models. The main contributions of this paper are summarized below: 
\begin{enumerate}
    \item We construct a knockoff-based procedure for FDR-controlled multiple testing when there are measurement errors and/or missing data in predictors. This procedure can work on general conditional dependence models $Y|\X$ and data structures in $\X$. It can also identify mutual signals for multiple outcomes (e.g. BC and CRC).
    \item We demonstrate the FDR control property and the power of our method with extensive simulation settings. We also illustrate the application with Women's Health Initiative (WHI) data examples.
\end{enumerate}

The rest of the paper is organized as follows. In Section \ref{sec:meth}, we present notations and details of our proposed variable selection framework to control FDR when there are missing data and measurement errors in the predictors. The method can also identify mutual signals for multiple outcomes. In Section \ref{sec:simulation}, we show the empirical performance of the proposed method under different model assumptions and data structures. Finally, in Section \ref{sec:application}, {\color{black}we apply the proposed method to a nested case-control study of BC and CRC among WHI Bone Mineral Density (BMD) Subcohort data.}


\section{Methods}
\label{sec:meth}

\subsection{Notation}
Denote the symbol $[\cdot]$ by $[N] = \{1, \cdots, N\}$. For the underlying data without missing data and measurement error, we assume it can be sampled from $ (Y,\X)$ with $Y\in \R$ being the response variable and $\X\in \R^{p}$ being the $p$ predictors. Assume we have $n$ samples with $(Y_{i}, X_{i1},\cdots,X_{ip}) \iidsim \mathcal{D}, ~\text{for}~ i \in [n]$. We work on the multiple testing problem on the null hypotheses 
$H_{0j}:= Y \indep X_{j}|\X_{-j}, ~\text{where}~ \X_{-j} := \X \setminus \X_{j} ~\text{for}~ j \in [p].$ We aim at developing a selection procedure returning a selection set $\widehat{\mathcal{S}} \subseteq [p]$ with a controlled FDR:  
\begin{equation}\label{eqn:sfdr}
\text{FDR}(\widehat{\mathcal{S}}) = \EE{\text{FDP}(\widehat{\mathcal{S}})}= \EE{\frac{|\widehat{\mathcal{S}} \cap \mathcal{H}|}{|\widehat{\mathcal{S}}|\vee 1}},
\end{equation}
where $\mathcal{H} = \{j\in [p]: H_{0j} ~\text{is true}\}$.

Given the potential measurement error, we assume $\X$ is not available and an error-prone version $\W \in \R^p$ is available, where $\W=\X+\bmepsilon_{w}$ with $\bmepsilon_{w}\iidsim \mathcal{F}_w$ where $\mathcal{F}_w$ can be estimated from external data sources. Typically, we can assume a multivariate normal distribution for $\bmepsilon_{w}$, i.e., $\bmepsilon_{w}\sim \mathcal{N}(\mathbf{0}, \mathbf{\Sigma}_{\epsilon})$. Also, due to the potential missing data, not all data is available, so we further introduce indicator variables $\mathbf{R} \in \R^p$ to indicate the missing mechanism. We let $R_{j}=1$ to indicate that $W_{j}$ is observable and we adopt the missing at random (MAR) assumption, i.e., $\mathbb{P}(R_{j}=1|\W,\X)=\mathbb{P}(R_{j}=1|\W_{-j})$. 
Notice that this assumption is slightly stronger than the usual assumption of MAR ($\mathbb{P}(R_{j}=1|\W,\X)=\mathbb{P}(R_{j}=1|\X_{-j})$ ) based on true variables due to the potential measurement error. {\color{black}
In metabolomic data, this assumption is reasonable, since the missingness is mostly related to the signal detected from the other similar metabolite peaks, rather than the true underlying concentration of other metabolites.} For the remaining of the paper, with $n$ observations, our final observed data will be denoted as $(\Y, \mathbf{R}, \mathbf{R}\odot \mathbf{W})$, where $\odot$ represents the Hadamard product. Here $\Y \in \R^n$ is a vector of responses for the $n$ individuals, $\W \in \R^{n \times p}$ denotes the matrix with elements $W_{ij}$ for individual $i$ and predictor $j$, $\mathbf{R} \in \R^{n \times p}$ denotes the matrix with elements $R_{ij}$ the indicator variable for individual $i$ and predictor $j$. We use $\mathbf{\Sigma}$ to denote the variance-covariance matrix of $\W$.

\subsection{Imputation of missing data}
In general, we can consider that we will generate $K$ imputed datasets, denoted as $(\Y, \mathbf{W}^k)$, for $k\in [K]$. When $K=1$, we can consider simple imputation by half min value or mean value depending on our assumption on whether the missing is random or is due to the detection limit. Also, we can consider the multiple imputation method with $K\geq 1$ where each dataset can be generated using a chained equation approach. Specifically, we will first randomly impute the missing values based on the marginal distribution estimated from those individuals with the variable observed, i.e., 
$
W_{ij}^{k}=R_{ij}W_{ij}+(1-R_{ij})\frac{\sum_{i}R_{ij} W_{ij}}{\sum_i R_{ij}}.
$
Then we will update the imputed value iteratively over all $j\in [p]$ that $\sum_i R_{ij}<n$. First, we will fit a regression model of $W_{ij}$ on $\W_{i,-j}^{k}$ and $Y_i$ for those $R_{ij}=1$. Then we will update $W_{ij}^{k}$ for those $R_{ij}=0$ using the predicted value from the model and current $\W_{i,-j}^{k}$, $Y_i$. Here the regression models can be in the form of generalized linear models (\textit{default}), classification and regression trees (\textit{cart}), or random forest (\textit{rf}). 
{\color{black}Alternatively, when a large unlabled subsample exists, another option is to perform multiple imputations excluding the outcome $\Y$ from the above steps. We explore both options numerically in Section \ref{sec:simulation}.} 

\subsection{Knockoff construction}
For each imputed dataset $\W^k$, we can construct the knockoff $\widetilde{\W}^k$ using second order Model-X knockoff by sampling $\widetilde{\W}^k$ from $\mathcal{N}(\widetilde{\mathbf{\mu}},\widetilde{\mathbf{\Sigma}})$, where $\widetilde{\mathbf{\mu}}=\W^k -\W^k \mathbf{\Sigma}^{-1} \diag\{\mathbf{s}\}$ and $\widetilde{\mathbf{\Sigma}} = 2\diag\{\mathbf{s}\} - \diag\{\mathbf{s}\} \mathbf{\Sigma}^{-1} \diag\{\mathbf{s}\}$, such that
\begin{equation*}
Cov([\W^k,\widetilde{\W^k}]) = \left(\begin{array}{cc}\mathbf{\Sigma}&\mathbf{\Sigma}-\diag\{\mathbf{s}\}\\\mathbf{\Sigma}-\diag\{\mathbf{s}\}&\mathbf{\Sigma}\end{array}\right).\end{equation*} 
where $\mathbf{s}$ satisfies $\widetilde{\mathbf{\Sigma}} = 2\diag\{\mathbf{s}\} - \diag\{\mathbf{s}\} \mathbf{\Sigma}^{-1} \diag\{\mathbf{s}\}$ is semi-positive definite and $\mathbf{s}$ can be solved using the approximate semidefinite program (ASDP) algorithm as given in \citep{candes2018}. We use the R function \textit{create.second} within the R package \textit{knockoff} to implement this construction method.

\subsection{Test statistics}
We consider the following types of test statistics:
\begin{itemize}
\item Absolute coefficient of Lasso: We assume a working model in a generalized linear model (GLM) framework
$  f_Y(y ; \theta, \phi)=\exp \left\{\frac{y \theta-b(\theta)}{a(\phi)}+c(y, \phi)\right\}
 $
  where $\theta=\mathbf{x}^\top \boldsymbol{\beta}$. The expected response is given by the mean function $\mu(\theta)=b^{\prime}(\theta)=g^{-1}(\theta)$, where $g^{-1}(\cdot)$ is the inverse of a canonical link function $g(\cdot)$. We choose the Lasso variable selection procedure and construct the statistics as $\widehat{\bmbeta}(\lambda)$, where 
$
\widehat{\bmbeta}(\lambda)=\arg\min_{\bmbeta\in \R^{2p}} \sum_{i=1}^{n}\frac{(Y_i-\bmbeta^{\top}[\W_i \widetilde{\W}_i])^2}{V_i}+\lambda \norm{\bmbeta}_1,  
$ $V_i=V(g^{-1}(\bmbeta^{\top}[\W_i \widetilde{\W}_i]))$ and $V(\cdot)$ is the variance function specified for the GLM for $\Y$. Then we use the absolute value $|\widehat{\beta}_j(\lambda)|$ as defined above with a specific $\lambda$ value or $\lambda$ selected from cross-validation as test statistics (i.e., $Z_j=|\widehat{\beta}_j(\lambda)|$ and $\widetilde{Z}_j=|\widehat{\beta}_{p+j}(\lambda)|$ for $j\in [p]$). 
\item Lasso Order: We assume the same GLM model and run over a range of $\lambda$ values decreasing from $+\infty$ (a fully sparse model) to $0$ (a fully dense model) and define $Z_j$ ($\widetilde{Z}_j$) as the maximum $\lambda$ such that $\widehat{\beta}_j(\lambda)\neq 0$ ($\widehat{\beta}_{p+j}(\lambda)\neq 0$). If there is no $\lambda$ such that $\widehat{\beta}_j(\lambda)\neq 0$ ($\widehat{\beta}_{p+j}(\lambda)\neq 0$), then we will simply define $Z_j$ ($\widetilde{Z}_j$) as 0. 
\item Random Forest: We can use the variable importance factors from the random forest fitting of $Y$ on $[\W\ \widetilde{\W}]$ with either fixed tuning parameters or tuning parameters selected from cross-validation as $[\begin{array}{cc}\Z &\widetilde{\Z}\end{array}]$ 
\item Absolute coefficient of Generalized Dantzig Selector (GDS): We choose GDS variable selection procedure and construct the statistics as $\widehat{\boldsymbol{\beta}}_{\mathrm{DS}}(\lambda)$, where\\
$\widehat{\boldsymbol{\beta}}_{\mathrm{DS}}(\lambda)=\underset{\boldsymbol{\beta}\in \R^{2p}}{\operatorname{argmin}}\left[\|\boldsymbol{\beta}\|_1: \max_{1 \leq j \leq 2p}\left|\frac{1}{n} \sum_{i=1}^n W_{i j}\widetilde{W}_{i j}\left\{Y_i-\mu\left(\mathbf{[W_i\widetilde{W_i}]}^\top \boldsymbol{\beta}\right)\right\}\right| \leq \lambda\right]$. 
  \item Absolute coefficient of Generalized of Matrix Uncertainty Selector (GMUS): The test statistics is a feasible solution of 
  \[
\widehat{\boldsymbol{\beta}}_{\mathrm{MU}}(\lambda)=\underset{\boldsymbol{\beta}\in \R^{2p}}{\operatorname{argmin}}\left\{\|\boldsymbol{\beta}\|_1: \frac{1}{n}\left\|\mathbf{[W\widetilde{W}]}^\top(\mathbf{Y}-\mathbf{[W\widetilde{W}]} \boldsymbol{\beta})\right\|_{\infty} \leq \lambda+\delta\|\boldsymbol{\beta}\|_1\right\}.
  \]
  \item Absolute coefficient of Corrected Lasso: The test statistics can be defined as minimizing the loss
  $
  \widehat{\boldsymbol{\beta}}_{\mathrm{RCL}}(d)=\underset{\boldsymbol{\beta}:\|\boldsymbol{\beta}\|_1 \leq d}{\arg \min }\left\{\sum_{i=1}^{n}\frac{(Y_i-\bmbeta^{\top}[\W_i \widetilde{\W}_i])^2}{V_i}-\bmbeta^{\top} \mathbf{\Sigma}_{\epsilon} \bmbeta\right\}
  $
  where $\mathbf{\Sigma}_{\epsilon}$ is the variance-covariance matrix for the measurement error in $\W$, and $d$ can be a pre-fixed tuning parameter or selected from cross-validation. 
\end{itemize}

\subsection{Variable selection}
We first compute the p-value for each feature based on the formula $
p_j=\frac{1+\sum_{k=1}^K \One{Z_j^k\leq \widetilde{Z}_j^k}}{1+K}.
$
Then we reorder the feature index by the decreasing order of $\max_{k=1}^K \max\{|Z_j^k|,|\widetilde{Z}_j^k|\}$
and apply the Selective SeqStep and Selective SeqStep+ procedure \citep{barber2015} with $p$-value threshold $0.5$ to find the selection threshold $\widehat{k}_c$ such that
\begin{eqnarray*}
    \widehat{k}_c=\max\left\{j\in [p]: \frac{c+\sum_{k=1}^j \One{p_k>1/2}}{\sum_{k=1}^j \One{p_k\leq 1/2}\vee 1}\leq q\right\},
\end{eqnarray*}
for $c=0,1$ where $q$ is the FDR level to be controlled at. Then the final selection sets will be $\widehat{S}_{c}=\{j:p_j<1/2\}\cap [\widehat{k}_{c}]$.

Notice that when the imputation model is correct, and there is no measurement error, i.e., $\W=\X$, then we have the imputed datasets $(\Y, \W^k)\sim \mathcal{D}$ for each $k$, and thus by the knockoff construction and test statistics, it has been shown that $\mathbb{P}(Z_j^k< \widetilde{Z}_j^k)=0.5$ for $j\in\mathcal{H}$ \citep{barber2015,candes2018}. Since $p_j$ is symmetrically distributed and $\EE{p_j}=\frac{1+K/2}{1+K}=\frac{1}{2}+\frac{1}{2(1+K)}>\frac{1}{2}$ for $j\in \mathcal{H}$, therefore 
$\widehat{\text{FDP}} = \frac{c+\sum_{k=1}^j \One{p_k>1/2}}{\sum_{k=1}^j \One{p_k\leq 1/2}\vee 1}$ is a conservative estimate for the FDP and thus the Selective SeqStep procedure controls FDR \citep{barber2015}. 
In practice, due to the potential model misspecification and finite sample performance, the imputed distribution might be slightly different from the true distribution, nonetheless, the robustness of the knockoff approach \citep{barber2020} ensures that the FDR inflation is small. When there are measurement errors, the Corrected Lasso and GMUS are known to be able to handle measurement errors appropriately for coefficient estimation and 
thus the knockoff approach can be applied. The performance of other test statistics is expected to be affected by measurement errors. The impact of the scales and correlations of measurement errors will be evaluated numerically (see Section \ref{sec:simulation}).

\subsection{Joint selection for multiple outcomes}
Now we consider the mutual signal identification problem. Assume we have data from $M$ independent experiments and denote $[M] = \{1,\cdots,M\}$. Within the $m$-th experiment, the underlying complete data without measurement errors are $(Y^m_{i},X^m_{i1},\cdots,X^m_{ip}) \iidsim \mathcal{D}_m$, $i=1,\cdots,n_m$. In our setting, the outcome variables $Y^1, Y^2$ represent the two different cancer outcomes BC and CRC. Define $H_{0j}^m$ as the null hypothesis indicating the $j$-th feature not being a signal in the $m$-th experiment (i.e. $X_j^m \indep Y^m | \X^m_{-j}$ where $\X^m_{-j} := \{X^m_1,\cdots,X^m_{p}\} \setminus X^m_j$), and denote $\mathcal{H}^m = \{j\in [p]: H_{0j}^m ~\text{is true}\}$, where $[p]:=\{1,\cdots,p\}$. Instead of testing the $H_{0j}^m$s, we are interested in testing the union null hypotheses $
    H_{0j}=\cup_{m=1}^M H_{0j}^m, ~~\text{for}~ j \in [p].
$

We define $\mathcal{S} = \{j\in [p]: H_{0j} ~\text{is false}\}$ and $\mathcal{H} = \mathcal{S}^c = \cup_{m=1}^M \mathcal{H}^m = \{j\in [p]: H_{0j} ~\text{is true}\}$.
We aim at developing a selection procedure returning a selection set $\widehat{\mathcal{S}} \subseteq [p]$ {\color{black}with a controlled FDR, as defined in \eqref{eqn:sfdr}.}
For this task, we can get test statistics $Z_j^{km}$ and $\widetilde{Z}_{j}^{km}$ for each $m$ separately first as in sections 2.1-2.4 and then compute the p-values as
$
p_j=\frac{1+\sum_{k=1}^K \One{\prod_{m=1}^M (Z_j^{km}- \widetilde{Z}_j^{km})\leq 0}}{1+K}.
$
Then we can use this new $p$ value to get $\widehat{k}_c$ and $\widehat{S}_c$ reordering the feature index by the decreasing order of $\max_{k=1}^K \prod_{m=1}^M|Z_j^{km}-\widetilde{Z}_j^{km}|$  (or $\sum_{k=1}^K\prod_{m=1}^M|Z_j^{km}-\widetilde{Z}_j^{km}|$) . This approach is valid under a similar argument as in \citep{dai2021multiple}.

\section{Simulation} 
\label{sec:simulation}

In this section, we perform extensive numerical experiments to understand the finite sample performance of the proposed methods in Section \ref{sec:meth}.

\subsection{Data generation and settings}

We generate data with various sample sizes $n$ and dimension $p$. We sample the underlying feature matrix $\X$ with various variation and correlation settings; and the outcome $\Y$ from a sparse logistic regression with varied effect sizes. Then we sample the measurement errors $\bmepsilon_w$ and construct features with measurement errors $\W=\X+\bmepsilon_w$ with different measurment error scales and correlations. Next we sample the missing data indicators $\mathbf{R}$ under the missing at random (MAR) mechanism with varied missing proportion $p_{mis}$.

We consider three different settings:
\begin{enumerate}
    \item \textbf{Data with only missing data but not measurement errors.} Under this setting, we compare the performance of the following methods: Lasso, Lasso Order, RF with multiple imputation. We considered either including or excluding the outcome $\Y$ when performing the imputations and consider three different imputation methods (\textit{default} method using generalized linear models, classification and regression tree (\textit{cart}), and random forest (\textit{rf}).
    \item \textbf{Data with only measurement errors but not missing data.} We compare the performance of the following methods: Lasso, Lasso Order, RF, GDS, GMUS, and Corrected Lasso.
    \item \textbf{Data with both missing data and measurement errors.} We compare the performance of the following methods: Lasso, Lasso Order, RF, GDS, GMUS, and Corrected Lasso with multiple imputations.
\end{enumerate}
More details on the data generation and simulation settings are postponed in Web Appendix \ref{APP:A.1}.




\subsection{Results}
For Setting 1 with only missing data, we compare the performance of three variable selection methods: Lasso, Lasso Order, and RF (Table \ref{tab:sim1}). 
Across the variety of settings we experimented, all three variable selection methods effectively control the FDR empirically. The Lasso variable selection method demonstrates the highest power among the tested methods, followed by Lasso Order, and finally, RF. The RF method tends to be somewhat conservative, resulting in marginally lower power, which may be attributed to the Lasso method possessing the correctly specified model. As the sample size ($n$) decreases and the number of variables ($p$) increases, all methods maintain satisfactory FDR control, albeit with a slight reduction in power, as anticipated. The proportion of missing data does not significantly impact the performance of these methods in terms of FDR and power. When examining the three imputation methods, their performances are found to be relatively similar. For smaller sample sizes ($n=400$), the \textit{default} method offers marginally better power, while the \textit{rf} method slightly outperforms the others for larger sample sizes. The decision to impute the dependent variable ($\Y$) does not substantially alter the performance of these methods.

In Table \ref{tab:sim2}, we present the FDP and power for our experiments conducted on data with measurement errors. We compare various variable selection methods, including Lasso, Lasso Order, RF, GDS, Corrected Lasso and GMUS. Notably, the Corrected Lasso and GMUS methods are specifically designed for measurement error correction. The results are organized according to the number of variables $p$, effect size, and scale. For FDP, Lasso, Lasso Order, RF, and GDS methods exhibit similar values across all scenarios. When $p$ is small and the effect size is large, the FDP is marginally higher than the nominal FDR. Corrected Lasso consistently displays lower FDP values than the nominal FDR, but it tends to be slightly over-conservative. The GMUS method generally yields FDP values comparable to those of Lasso and Lasso Order. Regarding power, the Lasso method persistently achieves higher values compared to other methods across all settings. The GDS and Lasso Order methods also demonstrate robust power, while the RF method consistently exhibits lower power. The GMUS method's power performance is comparable to that of Lasso Order and GDS, while Corrected Lasso consistently shows the lowest power values among all methods. Overall, as the number of variables increases, FDP experiences a slight decrease and power undergoes a more noticeable reduction. The effect size has a more pronounced impact on power, with larger effect sizes resulting in higher power values. The scale also affects power, with smaller scales generally leading to increased power values. However, the FDP values remain relatively stable irrespective of changes in effect size or scale.

Tables \ref{tab:sim3} summarize the FDR and power of our proposed methods under simulation settings with both measurement errors and missing data. The variable selection methods we included are the same as in Setting 2. Here we present the result where the missing probabilities depend on $\W$ rather than $\X$. The additional simulation results when the missing probabilities depend on $\X$ can be found in Web Appendix \ref{APP:A}. For the measurement errors we consider two levels of Scales. For imputation methods, we consider the default, cart, and rf. We compare the performance based on whether to impute $\Y$. The variable selection methods Lasso, Lasso Order, RF and GDS do not consider measurement errors, while Corrected Lasso and GMUS correct measurement errors. 

\begin{table}[H]
    \centering
    \caption{Simulation results (FDP and Power) for Setting 1 (missing data) varying $n$, $p$ and $p_{mis}$; with different imputation methods (Imp M) and choice of whether to include $\Y$ in the imputation (Imp Y).}
    \label{tab:sim1}
{\small
\begin{tabular}{ccccccccccc}
\hline
\multicolumn{5}{c}{} & \multicolumn{3}{c}{FDP} &  \multicolumn{3}{c}{Power} \\ \hline
$n$ & $p$ & $p_{mis}$ & Imp M & Imp Y & Lasso & Lasso Order & RF & Lasso &  Lasso Order &  RF \\ \hline
400 & 60 & 0.05 & cart     & yes & 0.15 & 0.14 & 0.14 & 0.99 & 0.65 & 0.46 \\ 
400 & 60 & 0.05 & cart     & no  & 0.16 & 0.15 & 0.14 & 0.98 & 0.67 & 0.50 \\ 
400 & 60 & 0.05 & default  & yes & 0.18 & 0.13 & 0.15 & 0.99 & 0.64 & 0.50 \\ 
400 & 60 & 0.05 & default  & no  & 0.18 & 0.13 & 0.16 & 0.99 & 0.66 & 0.52 \\ 
400 & 60 & 0.05 & rf       & yes & 0.17 & 0.14 & 0.14 & 0.99 & 0.66 & 0.50 \\ 
400 & 60 & 0.05 & rf       & no  & 0.16 & 0.15 & 0.13 & 0.99 & 0.65 & 0.49 \\ 
400 & 60 & 0.15 & cart     & yes & 0.17 & 0.14 & 0.17 & 0.98 & 0.64 & 0.54 \\ 
400 & 60 & 0.15 & cart     & no  & 0.15 & 0.14 & 0.13 & 0.98 & 0.67 & 0.47 \\ 
400 & 60 & 0.15 & default  & yes & 0.17 & 0.15 & 0.13 & 0.99 & 0.68 & 0.50 \\ 
400 & 60 & 0.15 & default  & no  & 0.17 & 0.17 & 0.14 & 0.99 & 0.70 & 0.49 \\ 
400 & 60 & 0.15 & rf       & yes & 0.17 & 0.15 & 0.15 & 0.98 & 0.67 & 0.50 \\ 
400 & 60 & 0.15 & rf       & no  & 0.18 & 0.17 & 0.16 & 0.98 & 0.70 & 0.52 \\ 
1000 & 60 & 0.05 & cart     & yes & 0.17 & 0.18 & 0.18 & 1.00 & 0.94 & 0.76 \\ 
1000 & 60 & 0.05 & cart     & no  & 0.17 & 0.19 & 0.15 & 1.00 & 0.95 & 0.72 \\ 
1000 & 60 & 0.05 & default  & yes & 0.20 & 0.17 & 0.16 & 1.00 & 0.93 & 0.70 \\ 
1000 & 60 & 0.05 & default  & no  & 0.18 & 0.19 & 0.16 & 1.00 & 0.94 & 0.73 \\ 
1000 & 60 & 0.05 & rf       & yes & 0.19 & 0.17 & 0.17 & 1.00 & 0.92 & 0.74 \\ 
1000 & 60 & 0.05 & rf       & no  & 0.18 & 0.17 & 0.17 & 1.00 & 0.93 & 0.73 \\ 
1000 & 60 & 0.15 & cart     & yes & 0.18 & 0.20 & 0.18 & 1.00 & 0.94 & 0.73 \\ 
1000 & 60 & 0.15 & cart     & no  & 0.19 & 0.20 & 0.18 & 1.00 & 0.94 & 0.74 \\ 
1000 & 60 & 0.15 & default  & yes & 0.20 & 0.18 & 0.16 & 1.00 & 0.94 & 0.73 \\ 
1000 & 60 & 0.15 & default  & no  & 0.19 & 0.18 & 0.18 & 1.00 & 0.93 & 0.73 \\ 
1000 & 60 & 0.15 & rf       & yes & 0.22 & 0.22 & 0.18 & 1.00 & 0.94 & 0.75 \\ 
1000 & 60 & 0.15 & rf       & no  & 0.21 & 0.21 & 0.18 & 1.00 & 0.93 & 0.74 \\ 
1000 & 120 & 0.05 & cart & yes & 0.18 & 0.17 & 0.17 & 1.00 & 0.85 & 0.58 \\

1000 & 120 & 0.05 & cart & no & 0.18 & 0.16 & 0.16 & 1.00 & 0.84 & 0.59 \\

1000 & 120 & 0.05 & default & yes & 0.19 & 0.15 & 0.16 & 1.00 & 0.84 & 0.57 \\

1000 & 120 & 0.05 & default & no & 0.19 & 0.16 & 0.15 & 1.00 & 0.84 & 0.56 \\

1000 & 120 & 0.05 & rf & yes & 0.18 & 0.16 & 0.16 & 1.00 & 0.85 & 0.58 \\

1000 & 120 & 0.05 & rf & no & 0.18 & 0.17 & 0.16 & 1.00 & 0.86 & 0.58 \\

1000 & 120 & 0.15 & cart & yes & 0.19 & 0.18 & 0.17 & 1.00 & 0.85 & 0.59 \\

1000 & 120 & 0.15 & cart & no & 0.17 & 0.18 & 0.17 & 1.00 & 0.85 & 0.58 \\

1000 & 120 & 0.15 & default & yes & 0.20 & 0.19 & 0.16 & 1.00 & 0.87 & 0.58 \\

1000 & 120 & 0.15 & default & no & 0.18 & 0.18 & 0.18 & 1.00 & 0.84 & 0.60 \\

1000 & 120 & 0.15 & rf & yes & 0.20 & 0.18 & 0.19 & 1.00 & 0.85 & 0.62 \\

1000 & 120 & 0.15 & rf & no & 0.19 & 0.19 & 0.18 & 1.00 & 0.85 & 0.59 \\
\hline 
\end{tabular}
}
\end{table}


\begin{table}[H]
    \centering
    \caption{Simulation results (FDP and Power) for Setting 2 (data with measurement errors) for $n=1000$, varying $p$, scales of the effect ($A_{\bmbeta}$) and errors ($\sigma_{\epsilon}^2$).}
    \label{tab:sim2}
   \begin{tabular}{ccccccccc}
\hline
\multicolumn{2}{c}{FDP} & & & \\ \hline
p & $A_{\bmbeta}$ & $\sigma_{\epsilon}^2$ & Lasso & Lasso Order & RF & GDS  & Corrected Lasso & GMUS \\ \hline
60  & 0.5    & 0.6   & 0.23  & 0.21        & 0.22 & 0.21  & 0.13& 0.21      \\ 
60  & 0.5    & 1     & 0.23  & 0.22        & 0.20 & 0.21 & 0.19 & 0.22       \\ 
60  & 1.5    & 0.6   & 0.25  & 0.25        & 0.24 & 0.22 & 0.15 & 0.22       \\ 
60  & 1.5    & 1     & 0.25  & 0.25        & 0.28 & 0.23 & 0.25 & 0.25      \\ 
120 & 0.5    & 0.6   & 0.21  & 0.21        & 0.17 & 0.20 & 0.05 & 0.20       \\ 
120 & 0.5    & 1     & 0.18  & 0.19        & 0.16 & 0.18 & 0.14 & 0.17       \\ 
120 & 1.5    & 0.6   & 0.20  & 0.20        & 0.17 & 0.18 & 0.07 & 0.17       \\ 
120 & 1.5    & 1     & 0.20  & 0.20        & 0.20 & 0.20 & 0.10  & 0.19     \\ \hline
\multicolumn{2}{c}{Power} & & & \\ \hline
p & $A_{\bmbeta}$ & $\sigma_{\epsilon}^2$ & Lasso & Lasso Order & RF & GDS & Corrected Lasso & GMUS  \\ \hline
60  & 0.5    & 0.6   & 0.91  & 0.84        & 0.71 & 0.91 & 0.44 & 0.89       \\ 
60  & 0.5    & 1     & 0.88  & 0.79        & 0.61 & 0.87 & 0.49 & 0.83       \\ 
60  & 1.5    & 0.6   & 0.94  & 0.88        & 0.75 & 0.94 & 0.44 & 0.92       \\ 
60  & 1.5    & 1     & 0.93  & 0.87        & 0.75 & 0.92 & 0.63 & 0.90       \\ 
120 & 0.5    & 0.6   & 0.80  & 0.72        & 0.49 & 0.78 & 0.17 & 0.76       \\ 
120 & 0.5    & 1     & 0.70  & 0.59        & 0.40 & 0.67 & 0.25 & 0.64       \\ 
120 & 1.5    & 0.6   & 0.83  & 0.77        & 0.54 & 0.81 & 0.15 & 0.80      \\ 
120 & 1.5    & 1     & 0.76  & 0.65        & 0.49 & 0.74 & 0.25 & 0.72       \\ \hline
\end{tabular}
\end{table}



\begin{table}[H]
\centering
\caption{Simulation results (FDP and Power) for Setting 3 (data with both measurement errors and missing data) for $n=1000$, $p=60$, $p_{mis}=0.15$, and $A_{\bmbeta}=1$, varying $\sigma_{\epsilon}^2$, Imp M and Imp Y.}
\label{tab:sim3}
\begin{tabular}{ccccccccc}
\hline
\multicolumn{2}{c}{FDP} & & & \\ \hline
$\sigma_{\epsilon}^2$ & Imp M & Imp Y & Lasso & Lasso Order & RF & GDS & Corrected Lasso & GMUS \\ \hline
0.1    & default & yes & 0.18 & 0.19 & 0.15 & {\color{black}0.20} & 0.02 & {\color{black}0.15} \\
0.1     & default  & no & 0.16 & 0.20 & 0.18 & {\color{black}0.19} & 0.02 & {\color{black}0.13} \\
0.1     & cart   & yes  & 0.16 & 0.20 & 0.18 & {\color{black}0.19} & 0.02 & {\color{black}0.13} \\
0.1     & cart  & no    & 0.18 & 0.18 & 0.17 & {\color{black}0.17} & 0.03 & {\color{black}0.12} \\
0.1     & rf   & yes    & 0.17 & 0.19 & 0.17 & {\color{black}0.19} & 0.02 & {\color{black}0.14} \\
0.1       & rf   & no   & 0.17 & 0.20 & 0.15 & {\color{black}0.19} & 0.02 & {\color{black}0.14} \\
0.6  & default & yes    & 0.20 & 0.25 & 0.25 & {\color{black}0.20} & 0.19 & {\color{black}0.17} \\
0.6  & default & no     & 0.18 & 0.26 & 0.27 & {\color{black}0.20} & 0.21 & {\color{black}0.17} \\
0.6  & cart & yes       & 0.20 & 0.26 & 0.26 & {\color{black}0.20} & 0.19 & {\color{black}0.18} \\
0.6  & cart & no        & 0.19 & 0.28 & 0.26 & {\color{black}0.21} & 0.18 & {\color{black}0.19} \\
0.6  & rf & yes         & 0.20 & 0.27 & 0.28 & {\color{black}0.21} & 0.20 & {\color{black}0.16} \\
0.6  & rf & no          & 0.19 & 0.28 & 0.26 & {\color{black}0.20} & 0.18 & {\color{black}0.18} \\
\hline
\multicolumn{2}{c}{Power} & & & \\ \hline
$\sigma_{\epsilon}^2$ & Imp M & Imp Y & Lasso & Lasso Order & RF & GDS & Corrected Lasso & GMUS \\ \hline
0.1  & default & yes  & 1.00 & 0.92 & 0.80 & {\color{black}1.00} & 0.46 & {\color{black}0.99} \\
0.1  & default & no   & 1.00 & 0.92 & 0.79 & {\color{black}1.00} & 0.42 & {\color{black}0.98} \\
0.1  & cart & yes     & 1.00 & 0.92 & 0.80 & {\color{black}1.00} & 0.44 & {\color{black}0.98} \\
0.1  & cart & no      & 1.00 & 0.91 & 0.80 & {\color{black}1.00} & 0.43 & {\color{black}0.98} \\
0.1  & rf & yes       & 1.00 & 0.91 & 0.80 & {\color{black}1.00} & 0.42 & {\color{black}0.98}\\
0.1  & rf & no        & 1.00 & 0.92 & 0.78 & {\color{black}1.00} & 0.42 & {\color{black}0.98} \\
0.6  & default  & yes & {\color{black}0.94} & 0.86 & 0.78 & {\color{black}0.93} & 0.61 & {\color{black}0.93} \\
0.6  & default & no   & 0.93 & 0.85 & 0.77 & {\color{black}0.93} & 0.61 & {\color{black}0.91} \\
0.6  & cart & yes     & {\color{black}0.94} & 0.86 & 0.78 & {\color{black}0.93} & 0.57 & {\color{black}0.92} \\
0.6  & cart & no      & 0.93 & 0.86 & 0.77 & {\color{black}0.93} & 0.56 & {\color{black}0.92} \\
0.6  & rf & yes       & {\color{black}0.93} & 0.86 & 0.78 & {\color{black}0.92} & 0.60 & {\color{black}0.92} \\
0.6  & rf & no        & 0.93 & 0.86 & 0.77 & {\color{black}0.93} & 0.58 & {\color{black}0.91} \\
\hline
\end{tabular}

\end{table}

With the small scale of measurement errors (Scale = 0.1), the FDR for most methods are controlled close to or below the nominal value of 0.2, with Corrected Lasso consistently having the lowest FDR values across all conditions. The results are observed for both Type W and Type X, and for different Imp M and Imp Y conditions. With bigger scale measurement errors, for Type W, the FDR is under control for most of the methods, while for Type X, most methods fail to control the FDR. With large scale of measurement errors (Scale = 0.6), Lasso Order and RF methods also fail to control the FDR for the Type W settings. In terms of power, the GDS and GMUS have the best performance, followed by Lasso and Lasso Order, Corrected Lasso does not have good power. 

{\color{black} In Table S2 of Web Appendix \ref{APP:A.3}, we show the performance of proposed methods in Settings 2 and 3 for detecting mutual signals from 2 datasets. The simultaneous knockoff method with our procedure to handle missing data and measurement errors (GMUS) still controls the FDR and achieves comparable power.}

\section{Real data analysis}\label{sec:application}
We analyzed serum and urine specimens from the WHI BMD data (181 CRC cases; 577 BC cases; 758 matched controls) using several metabolomics platforms. The details on how the matched samples were selected can be found in Web Appendix \ref{APP:B.2}. We applied global metabolomics platforms (gas chromatography–mass spectrometry (GC-MS) and nuclear magnetic resonance (NMR)) for profiling urine metabolites and targeted platforms for profiling serum metabolites. In serum, using liquid chromatography with tandem mass spectrometry (LC-MS/MS), we targeted water-soluble metabolites covering over 50 major metabolic pathways, and using the recently developed Lipidyzer platform, we detected about 900 lipids from 13 different classes (Cholesterol ester(CE), Ceramides(CER), Diacylglycerol(DAG), Dihydroceramides(DCER), Free fatty acids(FFA), Hexosylceramides(HCER), Lactosylceramide(LCER), Lysophosphatidylcholine(LPC), Lysophosphatidylethanolamine(LPE), Phosphatidylcholine(PC), Phosphatidylethanolamine(PE), Sphingomyelin(SM), Triacylglycerol(TAG)). Over 1500 metabolites were obtained from urine and serum samples using these four complementary analytical platforms. {\color{black}The proportion of missing data as well as the signal noise ratio for measurement errors are summarized in Table S3 in Web Appendix \ref{APP:B} where we can see that the GCMS and LCMS suffer most from the measurement error with some SNR less than one and GCMS has the most missing data.} We applied the knockoff method with missing and measurement errors to find the metabolite factors associated with BC, CRC and shared factors for both cancers using the proposed methods.


\subsection{Data preprocessing}
We first preprocessed the data from the four different platforms (NMR, GCMS, LC-MS, and lipidomic) to remove outliers, batch effects, and variables with excessive missing values (details see Web Appendix \ref{APP:B.2}). For LC-MS and lipidomic data, we considered both concentration and composition data which led to a total of 6 groups of metabolites and we analyze each group separately. The binary variable case/control of certain cancer served as our response variable. Here the preprocessed metabolites in non-quality control (QC) samples forms the predictor matrix $W\in \R^{n \times p}$, where $n$ is the number of patients and $p$ is the number of metabolites.  

To achieve the goals that finding the risk metabolite factors for BC and CRC, we analyzed the data including the BC cases vs all controls, and the CRC cases vs all controls respectively. Then the summary statistics were combined using the method described in Section 2.6 to identify the shared factors associated with both breast and CRC. 

To build the model, first, for each platform dataset, we imputed the data by different imputation methods including half-min imputation, and multiple imputations with predictive mean matching method. For the multiple imputations, we performed MICE with ($K=5$) using different analytic data for each outcome (i.e., BC cases + all controls for BC analysis and CRC + all controls for CRC analysis) were used to impute the missing values, and outcomes were included in the multiple imputation procedure. The variance-covariance matrix for the measurement errors were estimated using QC samples (details can be found on Web Appendix \ref{APP:B.3}). Then, we applied three preferred methods (Lasso, Lasso Order, GMUS) and three alternative methods (RF, GDS, Corrected Lasso) as described in Section 2 to generate test statistics. Then we applied our knockoff and simultaneous knockoff variable selection procedures with a target FDR level of 0.1. To make our results more robust, we performed stability selection by running different methods 100 times and recording the percentage of the replications each variable was selected. 

For sensitivity analysis, we performed the same analysis as above but using different analytical data for each outcome (i.e., BC cases + matched BC control for BC analysis and CRC cases + matched CRC control for CRC analysis).

\subsection{Results}
We presented the metabolites that were selected to be associated with BC, CRC, and both of these two cancers with $\geq 50\%$ of the replications in Tables \ref{tab:selectbc}, \ref{tab:selectcc} and \ref{tab:selectboth}, respectively when using the three preferred variable selection methods (i.e., Lasso, Lasso Order, and GMUS). The directions of the marginal associations are also indicated ($+$: positive, $-$: negative). The list of the metabolites that were selected to be associated with BC, and CRC with $\geq 10\%$ of the replications using all methods can be found in Web Appendix \ref{APP:C}.

\begin{table}[H]
\centering
\caption{Metabolites that are robustly (selected among $\geq$50\% of replications) associated with BC risks and the direction of their marginal association to the BC risks.}
\label{tab:selectbc}
\begin{adjustbox}{max width=\textwidth,max totalheight=\textheight,keepaspectratio}
\begin{tabular}{llll}
\hline
Method & \begin{tabular}[c]{@{}l@{}}Platform\end{tabular} & \begin{tabular}[c]{@{}l@{}}Half Min Imputation\end{tabular} & \begin{tabular}[c]{@{}l@{}}Multiple Imputation\end{tabular} \\
\hline
   Lasso& \begin{tabular}[c]{@{}l@{}}GCMS\end{tabular} &
  \begin{tabular}[c]{@{}l@{}}Alpha$-$ketoglutarate ({\color{black}76}\%)(-)\end{tabular} &
  \begin{tabular}[c]{@{}l@{}}\end{tabular} \\ \hline

Lasso&\begin{tabular}[c]{@{}l@{}}Lipidyzer \\ (composition)\end{tabular} &
  \begin{tabular}[c]{@{}l@{}}\end{tabular} &
  \begin{tabular}[c]{@{}l@{}}TAG 48:5(FA18:3) (80\%)(-)\\DAG 14:1/18:1 (78\%)(+)\end{tabular} \\\hline
   
Lasso&  \begin{tabular}[c]{@{}l@{}}Lipidyzer \\ (concentration)\end{tabular}&
  \begin{tabular}[c]{@{}l@{}}DAG 14:1/18:1 (97\%)(+)\end{tabular}&
  \begin{tabular}[c]{@{}l@{}}DAG 14:1/18:1 (97\%)(+)\\ TAG 48:5(FA18:3) (64\%)(-)\end{tabular} \\ \hline
Lasso Order&\begin{tabular}[c]{@{}l@{}}NMR\end{tabular} &
  \begin{tabular}[c]{@{}l@{}}N$-$methylnicotinic acid (59\%)(+)\end{tabular} &
  \begin{tabular}[c]{@{}l@{}}N$-$methylnicotinic acid (63\%)(+)\end{tabular} \\ \hline
  
{\color{black}Lasso Order}&\begin{tabular}[c]{@{}l@{}}{\color{black}GCMS}\end{tabular} &
  \begin{tabular}[c]{@{}l@{}}{\color{black}Alpha$-$ketoglutarate (54\%)(-)}\end{tabular} &
  \begin{tabular}[c]{@{}l@{}} \end{tabular} \\ \hline
  
  Lasso Order&\begin{tabular}[c]{@{}l@{}}LCMS \\ (AbsQuant)\end{tabular} &
  \begin{tabular}[c]{@{}l@{}}3HBA (67\%)(+)\\ Cystine (66\%)(-)\end{tabular} &
  \begin{tabular}[c]{@{}l@{}}  \end{tabular} \\ \hline
  
Lasso Order&\begin{tabular}[c]{@{}l@{}}Lipidyzer \\ (composition)\end{tabular} &
  \begin{tabular}[c]{@{}l@{}}TAG 47:0(FA15:0) (68\%)(-) \end{tabular} &
  \begin{tabular}[c]{@{}l@{}}TAG 48:5(FA18:3) (80\%)(-)\\DAG 14:1/18:1 (76\%)(+) \end{tabular} \\ \hline
GMUS &\begin{tabular}[c]{@{}l@{}}Lipidyzer \\ (composition)\end{tabular} &
  \begin{tabular}[c]{@{}l@{}}TAG 52:2(FA18:2) (66\%)(-)\end{tabular} &
  \begin{tabular}[c]{@{}l@{}}PC 16:0/18:2 (53\%)(+)\end{tabular} \\ \hline
GMUS &  \begin{tabular}[c]{@{}l@{}}Lipidyzer \\ (concentration)\end{tabular} &
  \begin{tabular}[c]{@{}l@{}}DAG 14:1/18:1 (63\%)(+)\end{tabular}&
  \begin{tabular}[c]{@{}l@{}}DAG 14:1/18:1 (93\%)(+)\end{tabular} \\
\hline
{\color{black}Corrected Lasso}&\begin{tabular}[c]{@{}l@{}}{\color{black}GCMS}\end{tabular} &
  \begin{tabular}[c]{@{}l@{}}{\color{black}Alpha$-$ketoglutarate (55\%)(-)}\end{tabular} &
  \begin{tabular}[c]{@{}l@{}}{\color{black}Alpha$-$ketoglutarate (73\%)(-)} \end{tabular} \\\hline
  
\end{tabular}
\end{adjustbox}
\end{table}
 
\begin{table}[H]
\centering
\caption{Metabolites that are robustly (selected among $\geq$50\% of replications) associated with CRC risks and the direction of their marginal association to the CRC risks.}
\label{tab:selectcc}
\begin{adjustbox}{max width=\textwidth,max totalheight=\textheight,keepaspectratio}
\begin{tabular}{llll}
\hline
Method & \begin{tabular}[c]{@{}l@{}}Platform\end{tabular} & \begin{tabular}[c]{@{}l@{}}Half Min Imputation\end{tabular} & \begin{tabular}[c]{@{}l@{}}Multiple Imputation\end{tabular} \\
\hline
Lasso &\begin{tabular}[c]{@{}l@{}}GCMS\end{tabular} &
  \begin{tabular}[c]{@{}l@{}}\end{tabular} &
  \begin{tabular}[c]{@{}l@{}}2,3$-$Dihydroxybutanoic acid ({\color{black}93}\%)(+)\end{tabular}\\ 
  \hline 
Lasso &\begin{tabular}[c]{@{}l@{}}LCMS \\ (AbsQuant)\end{tabular} &
  \begin{tabular}[c]{@{}l@{}}Glucose (84\%)(+)\\ Cystine (52\%)(-)\end{tabular} &
  \begin{tabular}[c]{@{}l@{}}Glucose (62\%)(+)\end{tabular}\\
\hline
Lasso &\begin{tabular}[c]{@{}l@{}}LCMS \\ (RelQuant)\end{tabular} &
  \begin{tabular}[c]{@{}l@{}}Glycerate (69\%)(+)\\ Adenosine (66\%)(-)\end{tabular}
&
  \begin{tabular}[c]{@{}l@{}}Adenosine (69\%)(-)\\Glycerate (67\%)(+)\end{tabular}\\
\hline
Lasso &\begin{tabular}[c]{@{}l@{}}Lipidyzer \\ (composition)\end{tabular} &
  \begin{tabular}[c]{@{}l@{}}TAG 48:5(FA18:2) (59\%)(+)\end{tabular} &
  \begin{tabular}[c]{@{}l@{}}\end{tabular}\\
\hline
Lasso Order &\begin{tabular}[c]{@{}l@{}} NMR \end{tabular}&
  \begin{tabular}[c]{@{}l@{}}N$-$methylnicotinic acid (76\%)(-) \end{tabular}&
  \begin{tabular}[c]{@{}l@{}}N$-$methylnicotinic acid (54\%)(-) \end{tabular}\\
\hline
Lasso Order &\begin{tabular}[c]{@{}l@{}} GCMS \end{tabular}&
  \begin{tabular}[c]{@{}l@{}}\end{tabular} &
  \begin{tabular}[c]{@{}l@{}}2,3$-$Dihydroxybutanoic acid ({\color{black}63}\%)(+)\end{tabular} \\
\hline
Lasso Order &\begin{tabular}[c]{@{}l@{}}LCMS \\ (AbsQuant)\end{tabular} &
  \begin{tabular}[c]{@{}l@{}}Cystine (100\%)(-)\\ 3HBA (75\%)(+)\end{tabular} &
  \begin{tabular}[c]{@{}l@{}}Cystine (94\%)(-)\end{tabular}  \\
\hline
Lasso Order &\begin{tabular}[c]{@{}l@{}}LCMS \\ (RelQuant)\end{tabular} &
  \begin{tabular}[c]{@{}l@{}}Malate (57\%)(+) \end{tabular}&
  \begin{tabular}[c]{@{}l@{}}\end{tabular} \\
\hline
Lasso Order &\begin{tabular}[c]{@{}l@{}}Lipidyzer \\ (composition)\end{tabular} &
  \begin{tabular}[c]{@{}l@{}}TAG 48:5(FA18:2) (60\%)(+)\\ \end{tabular} &
  \begin{tabular}[c]{@{}l@{}}\end{tabular} \\
\hline
Lasso Order &\begin{tabular}[c]{@{}l@{}}Lipidyzer \\ (concentration)\end{tabular} &
  \begin{tabular}[c]{@{}l@{}}TAG 47:2(FA14:0) (61\%)(-)\end{tabular} &
  \begin{tabular}[c]{@{}l@{}}\end{tabular}  \\
\hline
GMUS & \begin{tabular}[c]{@{}l@{}}LCMS \\ (AbsQuant)\end{tabular} &
  \begin{tabular}[c]{@{}l@{}}\end{tabular} &
  \begin{tabular}[c]{@{}l@{}}Histidine (69\%)(-)\end{tabular}  \\
\hline
GMUS &  \begin{tabular}[c]{@{}l@{}}Lipidyzer \\ (concentration)\end{tabular} &
  \begin{tabular}[c]{@{}l@{}}CER 16:0 (52\%)(+)\end{tabular} &
  \begin{tabular}[c]{@{}l@{}}CER 16:0 (66\%)(+)\end{tabular}  \\
\hline
\end{tabular}
\end{adjustbox}
\end{table}

\begin{table}[H]
\centering
\caption{Metabolites that are robustly (selected among $\geq$50\% of replications) associated with both BC and CRC risks and the direction of their marginal association to these two cancer risks.}
\label{tab:selectboth}
\begin{adjustbox}{max width=\textwidth,max totalheight=\textheight,keepaspectratio}
\begin{tabular}{llll}
\hline
Method & \begin{tabular}[c]{@{}l@{}}Platform\end{tabular} & \begin{tabular}[c]{@{}l@{}}Half Min Imputation\end{tabular} & \begin{tabular}[c]{@{}l@{}}Multiple Imputation\end{tabular} \\
\hline
 Lasso Order &\begin{tabular}[c]{@{}l@{}}  NMR \end{tabular}&
  \begin{tabular}[c]{@{}l@{}}  \end{tabular}&
  \begin{tabular}[c]{@{}l@{}} N$-$methylnicotinic acid (57\%)(B:+)(C:-)\end{tabular} \\ \hline

Lasso Order &\begin{tabular}[c]{@{}l@{}}LCMS \\ (AbsQuant)\end{tabular} &
  \begin{tabular}[c]{@{}l@{}}Cystine (89\%)(B:-)(C:-)\\ 3HBA (68\%)(B:+)(C:+)\end{tabular} &
  \begin{tabular}[c]{@{}l@{}}Cystine (99\%)(B:-)(C:-)\\ 3HBA (83\%)(B:+)(C:+)\\ Glutamic acid (78\%)(B:+)(C:+)\end{tabular} \\\hline
  
GMUS &\begin{tabular}[c]{@{}l@{}}LCMS \\ (AbsQuant)\end{tabular} &
  \begin{tabular}[c]{@{}l@{}}Choline (56\%)(B:+)(C:+)\end{tabular} &
  \begin{tabular}[c]{@{}l@{}}Choline (63\%)(B:+)(C:+)\end{tabular} \\ \hline
\end{tabular}
\end{adjustbox}
\end{table}

 Comparing the three variable selection methods, Lasso Order gives the most selections, followed by Lasso and GMUS. Across different methods, there are metabolites that are mutually selected by the different methods, for example, TAG 48:5(FA 18:3) and DAG 14:1/18:1; on the other hand, each method also selects some unique metabolites. Comparing the two imputation options, they produce relatively consistent results. Since we only select variables with high replications ($\geq 50\%$, with FDR controlled for every replication), we include selections from all the proposed variable selection methods as identified signals. Sensitivity analysis using single cancer type cases and their own matched controls is performed. More details on the analysis and the selected metabolites are presented in Web Appendix \ref{APP:B}. The variables selected are largely the same, although fewer variables are selected due to reduced sample sizes.



\section{Discussion}
\label{sec:discussion}

In conclusion, we find that appropriately handling missing data and measurement errors using the knockoff approach will control FDR at the targeted rate and gain power in terms of finding metabolites associated with BC and CRC risk. When general simultaneous knockoff methods are used, we find that appropriately handling missing data and measurement error using will control FDR for multiple outcomes at the targeted rate.

We identified a group of metabolites that are associated with either BC, CRC, or both cancers. The biomarker findings are largely consistent with the existing literature. For example, pentanedioic acid derivatives have been proposed as a potential agent for the treatment of BC \citep{zhang2022metabolic}. N-methyl nicotinic acid level in LC-ESI-MS has been reported to be positively associated with BC \citep{valko2021breast}. The increase of 3-hydroxybutyric acid (3HBA) level has been found as an indication of the increased fatty acid oxidation, a hallmark for cancer aggressiveness \citep{cappelletti2017metabolic}. For CRC, serum 2,3$-$dihydroxybutanoic acid has been reported as a biomarker \citep{loktionov2020biomarkers}. Glucose \citep{vulcan2017high}, glycerate \citep{ni2014metabonomics}, adenosine \citep{hata2023adar1}, N-methyl nicotinic acid, cystine \citep{miller2013homocysteine}, malate \citep{neitzel2020targeting}, histidine \citep{rothwell2023circulating} and CER (16:0) \citep{machala2019colon} have also been discovered to be associated with CRC in other independent studies. Choline has been reported to be positively associated with risks for both BC \citep{bae2014plasma} and CRC \citep{xu2008choline}.  The results confirm some findings of previous literature and also discover a few new potential metabolite biomarkers for future validation. 

A limitation of the current study is the small sample size for QCs which leads to large variations in the estimation of the variance-covariance matrix of measurement error. This could make the measurement error correction method vulnerable to potential misspecification of the measurement error distribution and be sensitive to the result of an outlier in 1 or 2 QC pairs. In the current application, we use second-order Model-X knockoff construction. When the variables $X$ and measurement errors both follow the multivariate normal distribution, the second-order condition is sufficient to guarantee the exchangeability of the whole distribution. However, when the variable is non-gaussian distributed, the higher order moment mismatching could lead to the difference in the distribution of $Z$ and $\widetilde{Z}$ for null variables, which will affect the FDP from the knockoff, especially the simultaneous knockoff procedure. When the measurement error is non-gaussian, Corrected Lasso will not be suitable and estimating the optimal error bound for GMUS will be challenging and requires a larger sample size for QCs. Further measurement error correction methods for both the estimation of the effect and the variable selection will be worth future research.  Another issue is that some metabolites are highly correlated to each other, which will make the knockoff feature very close to the original feature and thus lead to low power. {\color{black}Further method development for group variable selection (by treating metabolites from the same pathway as a group or treating highly correlated metabolites as a group) is worth further exploration but is beyond the scope of this paper.} In addition, the current analysis is based on considering each platform's data separately. In the future, methods need to be developed to handle multiple platform data together by solving the challenge of very different measurement scales and potential screening methods to reduce the number of features to allow powerful knockoff construction.

\section*{Acknowledgements}
This research is partly supported by the National Cancer Institute under grants R01 CA119171, CA277133, and P30 CA015704 and by the National Institute of General Medical Sciences under grant U54 GM115458. The WHI programs are funded by the National Heart, Lung, and Blood Institute, National Institutes of Health, U.S. Department of Health and Human Services through contracts, HHSN268201600018C, HHSN268201600001C, HHSN268201600002C, HHSN268201600003C, and HHSN268201600004C.

The authors acknowledge the following investigators in the WHI Program: Program Office: Jacques E. Rossouw, Shari Ludlam, Dale Burwen, Joan McGowan, Leslie Ford, and Nancy Geller, National Heart, Lung, and Blood Institute, Bethesda, Maryland; Clinical Coordinating Center, Women’s Health Initiative Clinical Coordinating Center: Garnet L. Anderson, Ross L. Prentice, Andrea Z. LaCroix, and Charles L. Kooperberg, Public Health Sciences, Fred Hutchinson Cancer Center, Seattle, Washington; Investigators and Academic Centers: JoAnn E. Manson, Brigham and Women’s Hospital, Harvard Medical School, Boston,Massachusetts; Barbara V. Howard, MedStar Health Research Institute/Howard University, Washington, DC; Marcia L. Stefanick, Stanford Prevention Research Center, Stanford, California; Rebecca Jackson, The Ohio State University, Columbus, Ohio; Cynthia A. Thomson, University of Arizona, Tucson/Phoenix, Arizona; Jean Wactawski-Wende, University at Buffalo, Buffalo, New York; Marian C. Limacher, University of Florida, Gainesville/Jacksonville, Florida; Robert M. Wallace, University of Iowa, Iowa City/ Davenport, Iowa; Lewis H. Kuller, University of Pittsburgh, Pittsburgh, Pennsylvania; and Sally A. Shumaker, Wake Forest University School of Medicine, Winston-Salem, North Carolina; Women’s Health Initiative Memory Study: Sally A. Shumaker, Wake Forest University School of Medicine,Winston-Salem, North Carolina. For a list of all the investigators who have contributed to WHI science, please visit:\\ https://www.whi.org/researchers/SitePages/WHI\%20Investigators.aspx.

Decisions concerning study design, data collection and analysis, interpretation of the results, the preparation of the manuscript, and the decision to submit the manuscript for publication resided with committees that comprised WHI investigators and included National Heart, Lung, and Blood Institute representatives. The contents of the paper are solely the responsibility of the authors.

\section*{Data Availability Statement}
The data that support the findings in this paper is not publicly available but could be requested through WHI in a collaborative mode as described on the Women's Health Initiative website (www.whi.org). 

\bibliography{miss}

\begin{thebibliography}{57}
\providecommand{\natexlab}[1]{#1}
\providecommand{\url}[1]{\texttt{#1}}
\expandafter\ifx\csname urlstyle\endcsname\relax
  \providecommand{\doi}[1]{doi: #1}\else
  \providecommand{\doi}{doi: \begingroup \urlstyle{rm}\Url}\fi

\bibitem[ACS(2020)]{american_cancer_society_2020}
ACS.
\newblock \emph{Cancer Facts and Figures 2020}.
\newblock American Cancer Society, Atlanta, GA, 2020.

\bibitem[Antoniadis et~al.(2010)Antoniadis, Fryzlewicz, Letué, and
  Sapatinas]{antoniadis2010}
Antoniadis, A., Fryzlewicz, P., Letué, F., and Sapatinas, T.
\newblock The dantzig selector in cox's proportional hazards model.
\newblock \emph{Scandinavian Journal of Statistics}, 37\penalty0 (4):\penalty0
  531--552, 2010.

\bibitem[Bae et~al.(2014)Bae, Ulrich, Neuhouser, Malysheva, Bailey, Xiao,
  Brown, Cushing-Haugen, Zheng, Cheng, et~al.]{bae2014plasma}
Bae, S., Ulrich, C.~M., Neuhouser, M.~L., Malysheva, O., Bailey, L.~B., Xiao,
  L., Brown, E.~C., Cushing-Haugen, K.~L., Zheng, Y., Cheng, T.-Y.~D., et~al.
\newblock Plasma choline metabolites and colorectal cancer risk in the women's
  health initiative observational study.
\newblock \emph{Cancer research}, 74\penalty0 (24):\penalty0 7442--7452, 2014.

\bibitem[Barber \& Candès(2015)Barber and Candès]{barber2015}
Barber, R.~F. and Candès, E.~J.
\newblock Controlling the false discovery rate via knockoffs.
\newblock \emph{Ann. Statist.}, 43\penalty0 (5):\penalty0 2055--2085, 2015.
\newblock \doi{10.1214/15-AOS1337}.

\bibitem[Barber \& Candès(2019)Barber and Candès]{barber2019}
Barber, R.~F. and Candès, E.~J.
\newblock A knockoff filter for high-dimensional selective inference.
\newblock \emph{Ann. Statist.}, 47\penalty0 (5):\penalty0 2504--2537, 2019.
\newblock \doi{10.1214/18-AOS1755}.

\bibitem[Barber et~al.(2020)Barber, Candès, and Samworth]{barber2020}
Barber, R.~F., Candès, E.~J., and Samworth, R.~J.
\newblock Robust inference with knockoffs.
\newblock \emph{Ann. Statist.}, 48\penalty0 (3):\penalty0 1409--1431, 2020.
\newblock \doi{10.1214/19-AOS1852}.

\bibitem[Bates et~al.(2020)Bates, Candès, Janson, and Wang]{bates2020}
Bates, S., Candès, E., Janson, L., and Wang, W.
\newblock Metropolized knockoff sampling.
\newblock \emph{Journal of the American Statistical Association}, 2020.
\newblock \doi{10.1080/01621459.2020.1729163}.

\bibitem[Benjamini \& Hochberg(1995)Benjamini and Hochberg]{benjamini1995}
Benjamini, Y. and Hochberg, Y.
\newblock Controlling the false discovery rate: A practical and powerful
  approach to multiple testing.
\newblock \emph{Journal of the Royal Statistical Society. Series B
  (Methodological)}, 57\penalty0 (1):\penalty0 289--300, 1995.
\newblock ISSN 00359246.

\bibitem[Bergstralh et~al.(1995)Bergstralh, Kosanke, et~al.]{bergstrahl}
Bergstralh, E.~J., Kosanke, J.~L., et~al.
\newblock Computerized matching of cases to controls.
\newblock Technical report, Technical report, 1995.

\bibitem[Bogomolov \& Heller(2013)Bogomolov and Heller]{bogomolov2013}
Bogomolov, M. and Heller, R.
\newblock Discovering findings that replicate from a primary study of high
  dimension to a follow-up study.
\newblock \emph{Journal of the American Statistical Association}, 108\penalty0
  (504):\penalty0 1480--1492, 2013.
\newblock \doi{10.1080/01621459.2013.829002}.

\bibitem[Bogomolov \& Heller(2018)Bogomolov and Heller]{bogomolov2018}
Bogomolov, M. and Heller, R.
\newblock {Assessing replicability of findings across two studies of multiple
  features}.
\newblock \emph{Biometrika}, 105\penalty0 (3):\penalty0 505--516, 2018.
\newblock ISSN 0006-3444.
\newblock \doi{10.1093/biomet/asy029}.

\bibitem[Candès et~al.(2018)Candès, Fan, Janson, and Lv]{candes2018}
Candès, E., Fan, Y., Janson, L., and Lv, J.
\newblock Panning for gold: ‘model-x’ knockoffs for high dimensional
  controlled variable selection.
\newblock \emph{Journal of the Royal Statistical Society: Series B (Statistical
  Methodology)}, 80\penalty0 (3):\penalty0 551--577, 2018.
\newblock \doi{https://doi.org/10.1111/rssb.12265}.

\bibitem[Cappelletti et~al.(2017)Cappelletti, Iorio, Miodini, Silvestri, Dugo,
  and Daidone]{cappelletti2017metabolic}
Cappelletti, V., Iorio, E., Miodini, P., Silvestri, M., Dugo, M., and Daidone,
  M.~G.
\newblock Metabolic footprints and molecular subtypes in breast cancer.
\newblock \emph{Disease markers}, 2017, 2017.

\bibitem[Chen et~al.(2019)Chen, Hou, and Hou]{chen2020}
Chen, J., Hou, A., and Hou, T.~Y.
\newblock {A prototype knockoff filter for group selection with FDR control}.
\newblock \emph{Information and Inference: A Journal of the IMA}, 9\penalty0
  (2):\penalty0 271--288, 2019.
\newblock ISSN 2049-8772.
\newblock \doi{10.1093/imaiai/iaz012}.

\bibitem[Cheung et~al.(2019)Cheung, Ma, Tse, Yeung, Tsang, Chu, Kan, and
  Cho]{cheung2019}
Cheung, P.-K., Ma, M.-H., Tse, H.-F., Yeung, K.-Y., Tsang, H.-C., Chu, M.-K.,
  Kan, C.-M., and Cho, W.~C.
\newblock The applications of metabolomics in the molecular diagnostics of
  cancer.
\newblock \emph{Expert Review of Molecular Diagnostics}, 19\penalty0
  (9):\penalty0 785--793, 2019.

\bibitem[Chi(2008)]{chi2008}
Chi, Z.
\newblock False discovery rate control with multivariate p -values.
\newblock \emph{Electron. J. Statist.}, 2:\penalty0 368--411, 2008.
\newblock \doi{10.1214/07-EJS147}.

\bibitem[Dai \& Barber(2016)Dai and Barber]{dai2016}
Dai, R. and Barber, R.
\newblock The knockoff filter for fdr control in group-sparse and multitask
  regression.
\newblock In Balcan, M.~F. and Weinberger, K.~Q. (eds.), \emph{Proceedings of
  The 33rd International Conference on Machine Learning}, volume~48 of
  \emph{Proceedings of Machine Learning Research}, pp.\  1851--1859, New York,
  New York, USA, 2016. PMLR.

\bibitem[Dai \& Zheng(2023)Dai and Zheng]{dai2021multiple}
Dai, R. and Zheng, C.
\newblock False discovery rate-controlled multiple testing for union null
  hypotheses: a knockoff-based approach.
\newblock \emph{Biometrics}, n/a\penalty0 (n/a), 2023.
\newblock \doi{https://doi.org/10.1111/biom.13848}.
\newblock URL \url{https://onlinelibrary.wiley.com/doi/abs/10.1111/biom.13848}.

\bibitem[Datta \& Zou(2017)Datta and Zou]{datta2017}
Datta, A. and Zou, H.
\newblock Cocolasso for high-dimensional error-in-variables regression.
\newblock \emph{Annals of Statistics}, 45:\penalty0 2400--2426, 2017.

\bibitem[Garcia et~al.(2009)Garcia, Ibrahim, and Zhu]{garcia2009}
Garcia, R.~I., Ibrahim, J.~G., and Zhu, H.
\newblock Variable selection in the cox regression model with covariates
  missing at random.
\newblock \emph{Biometrics}, 66:\penalty0 97, 2009.

\bibitem[Hata et~al.(2023)Hata, Shigeyasu, Umeda, Yano, Takeda, Yoshida, Fuji,
  Yoshida, Yasui, Umeda, et~al.]{hata2023adar1}
Hata, N., Shigeyasu, K., Umeda, Y., Yano, S., Takeda, S., Yoshida, K., Fuji,
  T., Yoshida, R., Yasui, K., Umeda, H., et~al.
\newblock Adar1 is a promising risk stratification biomarker of remnant liver
  recurrence after hepatic metastasectomy for colorectal cancer.
\newblock \emph{Scientific reports}, 13\penalty0 (1):\penalty0 2078, 2023.

\bibitem[Heller \& Yekutieli(2014)Heller and Yekutieli]{heller2014b}
Heller, R. and Yekutieli, D.
\newblock Replicability analysis for genome-wide association studies.
\newblock \emph{Ann. Appl. Stat.}, 8\penalty0 (1):\penalty0 481--498, 2014.
\newblock \doi{10.1214/13-AOAS697}.

\bibitem[Heller et~al.(2014)Heller, Bogomolov, and Benjamini]{heller2014}
Heller, R., Bogomolov, M., and Benjamini, Y.
\newblock Deciding whether follow-up studies have replicated findings in a
  preliminary large-scale omics study.
\newblock \emph{Proceedings of the National Academy of Sciences}, 111\penalty0
  (46):\penalty0 16262--16267, 2014.
\newblock ISSN 0027-8424.
\newblock \doi{10.1073/pnas.1314814111}.

\bibitem[His et~al.(2019)His, Viallon, Dossus, Gicquiau, Achaintre, Scalbert,
  Ferrari, Romieu, Onland-Moret, Weiderpass, et~al.]{his2019}
His, M., Viallon, V., Dossus, L., Gicquiau, A., Achaintre, D., Scalbert, A.,
  Ferrari, P., Romieu, I., Onland-Moret, N.~C., Weiderpass, E., et~al.
\newblock Prospective analysis of circulating metabolites and breast cancer in
  epic.
\newblock \emph{BMC Medicine}, 17\penalty0 (1):\penalty0 178, 2019.

\bibitem[Huang \& Janson(2020)Huang and Janson]{huang}
Huang, D. and Janson, L.
\newblock Relaxing the assumptions of knockoffs by conditioning.
\newblock \emph{Ann. Statist.}, 48\penalty0 (5):\penalty0 3021--3042, 2020.
\newblock \doi{10.1214/19-AOS1920}.

\bibitem[Johnson(2008)]{johnson2008}
Johnson, B.~A.
\newblock Variable selection in semiparametric linear regression with censored
  data.
\newblock \emph{Journal of the Royal Statistical Society: Series B (Statistical
  Methodology)}, 70:\penalty0 351, 2008.

\bibitem[Li et~al.(2021)Li, Sesia, Romano, Candès, and Sabatti]{li2021}
Li, S., Sesia, M., Romano, Y., Candès, E., and Sabatti, C.
\newblock {Searching for robust associations with a multi-environment knockoff
  filter}.
\newblock \emph{Biometrika}, 109\penalty0 (3):\penalty0 611--629, 11 2021.
\newblock ISSN 1464-3510.
\newblock \doi{10.1093/biomet/asab055}.
\newblock URL \url{https://doi.org/10.1093/biomet/asab055}.

\bibitem[Little \& Rubin(2002)Little and Rubin]{little2002}
Little, R.~J. and Rubin, D.~B.
\newblock \emph{Statistical Analysis with Missing Data}.
\newblock John Wiley \& Sons, New York, 2002.

\bibitem[Liu \& Zheng(2019)Liu and Zheng]{liu2019}
Liu, Y. and Zheng, C.
\newblock Deep latent variable models for generating knockoffs.
\newblock \emph{Stat}, 8\penalty0 (1):\penalty0 e260, 2019.
\newblock \doi{https://doi.org/10.1002/sta4.260}.
\newblock e260 sta4.260.

\bibitem[Loh \& Wainwright(2012)Loh and Wainwright]{loh2012}
Loh, P.-L. and Wainwright, M.~J.
\newblock {High-dimensional regression with noisy and missing data: Provable
  guarantees with nonconvexity}.
\newblock \emph{The Annals of Statistics}, 40\penalty0 (3):\penalty0 1637 --
  1664, 2012.
\newblock \doi{10.1214/12-AOS1018}.
\newblock URL \url{https://doi.org/10.1214/12-AOS1018}.

\bibitem[Loktionov(2020)]{loktionov2020biomarkers}
Loktionov, A.
\newblock Biomarkers for detecting colorectal cancer non-invasively: Dna, rna
  or proteins?
\newblock \emph{World journal of gastrointestinal oncology}, 12\penalty0
  (2):\penalty0 124, 2020.

\bibitem[Machala et~al.(2019)Machala, Proch{\'a}zkov{\'a}, Hofmanov{\'a},
  Kr{\'a}likov{\'a}, Slav{\'\i}k, Tylichov{\'a}, Ovesn{\'a}, Kozub{\'\i}k, and
  Vondr{\'a}{\v{c}}ek]{machala2019colon}
Machala, M., Proch{\'a}zkov{\'a}, J., Hofmanov{\'a}, J., Kr{\'a}likov{\'a}, L.,
  Slav{\'\i}k, J., Tylichov{\'a}, Z., Ovesn{\'a}, P., Kozub{\'\i}k, A., and
  Vondr{\'a}{\v{c}}ek, J.
\newblock Colon cancer and perturbations of the sphingolipid metabolism.
\newblock \emph{International journal of molecular sciences}, 20\penalty0
  (23):\penalty0 6051, 2019.

\bibitem[Miller et~al.(2013)Miller, Beresford, Neuhouser, Cheng, Song, Brown,
  Zheng, Rodriguez, Green, and Ulrich]{miller2013homocysteine}
Miller, J.~W., Beresford, S.~A., Neuhouser, M.~L., Cheng, T.-Y.~D., Song, X.,
  Brown, E.~C., Zheng, Y., Rodriguez, B., Green, R., and Ulrich, C.~M.
\newblock Homocysteine, cysteine, and risk of incident colorectal cancer in the
  women’s health initiative observational cohort.
\newblock \emph{The American journal of clinical nutrition}, 97\penalty0
  (4):\penalty0 827--834, 2013.

\bibitem[Nannini et~al.(2020)Nannini, Meoni, Amedei, and Tenori]{nannini2020}
Nannini, G., Meoni, G., Amedei, A., and Tenori, L.
\newblock Metabolomics profile in gastrointestinal cancers: Update and future
  perspectives.
\newblock \emph{World Journal of Gastroenterology}, 26\penalty0 (20):\penalty0
  2514--2532, 2020.

\bibitem[Neitzel et~al.(2020)Neitzel, Demuth, Wittmann, and
  Fahrer]{neitzel2020targeting}
Neitzel, C., Demuth, P., Wittmann, S., and Fahrer, J.
\newblock Targeting altered energy metabolism in colorectal cancer: oncogenic
  reprogramming, the central role of the tca cycle and therapeutic
  opportunities.
\newblock \emph{Cancers}, 12\penalty0 (7):\penalty0 1731, 2020.

\bibitem[Ni et~al.(2014)Ni, Xie, and Jia]{ni2014metabonomics}
Ni, Y., Xie, G., and Jia, W.
\newblock Metabonomics of human colorectal cancer: new approaches for early
  diagnosis and biomarker discovery.
\newblock \emph{Journal of proteome research}, 13\penalty0 (9):\penalty0
  3857--3870, 2014.

\bibitem[Playdon et~al.(2017)Playdon, Ziegler, Sampson, Stolzenberg-Solomon,
  Thompson, Irwin, Mayne, Hoover, and Moore]{playdon2017}
Playdon, M.~C., Ziegler, R.~G., Sampson, J.~N., Stolzenberg-Solomon, R.,
  Thompson, H.~J., Irwin, M.~L., Mayne, S.~T., Hoover, R.~N., and Moore, S.~C.
\newblock Nutritional metabolomics and breast cancer risk in a prospective
  study.
\newblock \emph{The American Journal of Clinical Nutrition}, 106\penalty0
  (2):\penalty0 637--649, 2017.

\bibitem[Putri et~al.(2013)Putri, Nakayama, Matsuda, and et~al.]{putri2013}
Putri, S.~P., Nakayama, Y., Matsuda, F., and et~al.
\newblock Current metabolomics: Practical applications.
\newblock \emph{Journal of Bioscience and Bioengineering}, 115\penalty0
  (6):\penalty0 579--589, 2013.

\bibitem[R\"{a}ssler et~al.(2013)R\"{a}ssler, Rubin, and Zell]{rassler2013}
R\"{a}ssler, S., Rubin, D.~B., and Zell, E.~R.
\newblock Imputation.
\newblock \emph{Wiley Interdisciplinary Reviews: Computational Statistics},
  5:\penalty0 20, 2013.

\bibitem[Romano et~al.(2020)Romano, Sesia, and Candès]{romano2019}
Romano, Y., Sesia, M., and Candès, E.
\newblock Deep knockoffs.
\newblock \emph{Journal of the American Statistical Association}, 115\penalty0
  (532):\penalty0 1861--1872, 2020.
\newblock \doi{10.1080/01621459.2019.1660174}.

\bibitem[Rosenbaum \& Tsybakov(2013)Rosenbaum and Tsybakov]{rosenbaum2013}
Rosenbaum, M. and Tsybakov, A.~B.
\newblock Improved matrix uncertainty selector.
\newblock In \emph{From Probability to Statistics and Back: High-Dimensional
  Models and Processes -- A Festschrift in Honor of Jon A. Wellner}, pp.\
  276--290, Beachwood, Ohio, USA, 2013. Institute of Mathematical Statistics.

\bibitem[Rothwell et~al.(2023)Rothwell, Be{\v{s}}evi{\'c}, Dimou, Breeur,
  Murphy, Jenab, Wedekind, Viallon, Ferrari, Achaintre,
  et~al.]{rothwell2023circulating}
Rothwell, J.~A., Be{\v{s}}evi{\'c}, J., Dimou, N., Breeur, M., Murphy, N.,
  Jenab, M., Wedekind, R., Viallon, V., Ferrari, P., Achaintre, D., et~al.
\newblock Circulating amino acid levels and colorectal cancer risk in the
  european prospective investigation into cancer and nutrition and uk biobank
  cohorts.
\newblock \emph{BMC medicine}, 21\penalty0 (1):\penalty0 1--13, 2023.

\bibitem[Sorensen et~al.(2015)Sorensen, Frigessi, Thoresen, and
  Glad]{sorensen2015}
Sorensen, O., Frigessi, A., Thoresen, M., and Glad, I.~K.
\newblock Measurement error in lasso: Impact and likelihood bias correction.
\newblock \emph{Statistica Sinica}, 25\penalty0 (2):\penalty0 809--829, 2015.

\bibitem[Spector \& Janson(2020)Spector and Janson]{spector2020}
Spector, A. and Janson, L.
\newblock Powerful knockoffs via minimizing reconstructability, 2020.

\bibitem[Tsiatis(2006)]{tsiatis2006}
Tsiatis, A.~A.
\newblock \emph{Semiparametric Theory and Missing Data}.
\newblock Springer, 2006.

\bibitem[Valko-Rokytovsk{\'a} et~al.(2021)Valko-Rokytovsk{\'a},
  O{\v{c}}en{\'a}{\v{s}}, Salayov{\'a}, and Kosteck{\'a}]{valko2021breast}
Valko-Rokytovsk{\'a}, M., O{\v{c}}en{\'a}{\v{s}}, P., Salayov{\'a}, A., and
  Kosteck{\'a}, Z.
\newblock Breast cancer: targeting of steroid hormones in cancerogenesis and
  diagnostics.
\newblock \emph{International Journal of Molecular Sciences}, 22\penalty0
  (11):\penalty0 5878, 2021.

\bibitem[Vulcan et~al.(2017)Vulcan, Manjer, and Ohlsson]{vulcan2017high}
Vulcan, A., Manjer, J., and Ohlsson, B.
\newblock High blood glucose levels are associated with higher risk of colon
  cancer in men: a cohort study.
\newblock \emph{BMC cancer}, 17\penalty0 (1):\penalty0 1--8, 2017.

\bibitem[WCRF(2018)]{wcrf_aicr_2018}
WCRF, A.
\newblock \emph{Continuous Update Project Expert Report: Diet, Nutrition,
  Physical Activity and Cancer: a Global Perspective}.
\newblock 2018.

\bibitem[Wolfson(2011)]{wolfson2011}
Wolfson, J.
\newblock Eeboost: A general method for prediction and variable selection based
  on estimating equations.
\newblock \emph{Journal of the American Statistical Association}, 106, 2011.

\bibitem[Xiao et~al.(2019)Xiao, Xia, Li, and et~al.]{xiao2019}
Xiao, Y., Xia, J., Li, L., and et~al.
\newblock Associations between dietary patterns and the risk of breast cancer:
  a systematic review and meta-analysis of observational studies.
\newblock \emph{Breast Cancer Research}, 21\penalty0 (1):\penalty0 16, 2019.

\bibitem[Xu et~al.(2008)Xu, Gammon, Zeisel, Lee, Wetmur, Teitelbaum, Bradshaw,
  Neugut, Santella, and Chen]{xu2008choline}
Xu, X., Gammon, M.~D., Zeisel, S.~H., Lee, Y.~L., Wetmur, J.~G., Teitelbaum,
  S.~L., Bradshaw, P.~T., Neugut, A.~I., Santella, R.~M., and Chen, J.
\newblock Choline metabolism and risk of breast cancer in a population-based
  study.
\newblock \emph{The FASEB journal: official publication of the Federation of
  American Societies for Experimental Biology}, 22\penalty0 (6):\penalty0 2045,
  2008.

\bibitem[Yang et~al.(2020)Yang, Wang, Cai, Wang, Shen, and Ke]{yang2020}
Yang, L., Wang, Y., Cai, H., Wang, S., Shen, Y., and Ke, C.
\newblock Application of metabolomics in the diagnosis of breast cancer: a
  systematic review.
\newblock \emph{Journal of Cancer}, 11\penalty0 (9):\penalty0 2540--2551, 2020.

\bibitem[Yusof et~al.(2012)Yusof, Isa, and Shah]{yusof2012}
Yusof, A.~S., Isa, Z.~M., and Shah, S.~A.
\newblock Dietary patterns and risk of colorectal cancer: a systematic review
  of cohort studies (2000-2011).
\newblock \emph{Asian Pacific Journal of Cancer Prevention}, 13\penalty0
  (9):\penalty0 4713--4717, 2012.

\bibitem[Zhang et~al.(2022)Zhang, Quinones, and Le]{zhang2022metabolic}
Zhang, C., Quinones, A., and Le, A.
\newblock Metabolic reservoir cycles in cancer.
\newblock In \emph{Seminars in cancer biology}. Elsevier, 2022.

\bibitem[Zhao \& Nguyen(2020)Zhao and Nguyen]{zhao2020}
Zhao, S.~D. and Nguyen, Y.~T.
\newblock Nonparametric false discovery rate control for identifying
  simultaneous signals.
\newblock \emph{Electron. J. Statist.}, 14\penalty0 (1):\penalty0 110--142,
  2020.
\newblock \doi{10.1214/19-EJS1663}.

\bibitem[Zheng et~al.(2021)Zheng, Gowda, Raftery, Neuhouser, Tinker, Prentice,
  Beresford, Zhang, Bettcher, Pepin, Djukovic, Gu, Barding, Song, and
  Lampe]{zheng2021evaluation}
Zheng, C., Gowda, G., Raftery, D., Neuhouser, M., Tinker, L., Prentice, R.,
  Beresford, S., Zhang, Y., Bettcher, L., Pepin, R., Djukovic, D., Gu, H.,
  Barding, G., Song, X., and Lampe, J.
\newblock Evaluation of potential metabolomic-based biomarkers of protein,
  carbohydrate and fat intakes using a controlled feeding study.
\newblock \emph{European Journal of Nutrition}, 60\penalty0 (8):\penalty0
  4207--4218, December 2021.
\newblock ISSN 1436-6207.
\newblock \doi{10.1007/s00394-021-02577-1}.

\bibitem[Zhu et~al.(2014)Zhu, Djukovic, Deng, Gu, Himmati, Chiorean, and
  Raftery]{zhu2014}
Zhu, J., Djukovic, D., Deng, L., Gu, H., Himmati, F., Chiorean, E., and
  Raftery, D.
\newblock Colorectal cancer detection using targeted serum metabolic profiling.
\newblock \emph{J. Proteome Res.}, 2014.

\end{thebibliography}
\bibliographystyle{icml2022}

\newpage
\appendix
\onecolumn
\section{Web Appendix A: Additional simulation details, settings and results}\label{APP:A}
\subsection{Web Appendix A.1: Simulation details} \label{APP:A.1}
 Denoting \label{sec:appendixA1} the sample size as $n$ and number of features $p$, we generate data with the combinations $(n,p)=(1000,60)$, $(1000,120)$ or $(400,60)$. 

First sample the feature matrix $\X \sim \mathcal{N}( \mathbf{0}, \mathbf{\Sigma}_X)$. Here we let $\mathbf{\Sigma}_X=AR(\sigma_X,\rho_X,p)$, where $AR(\sigma,\rho,p)$ denotes a $p\times p$ matrix with $(i,j)$-th element equals to $\sigma^2\rho^{|i-j|}$. We use $\sigma^2$ to control the magnitude of the variation, and $\rho$ to control the correlation between the predictors. Specifically, we set $\sigma_X^2=1$ and $\rho_X=0.5$. 

Second, we sample outcome $\Y$ from a logistic regression model
\begin{eqnarray*}
\log \frac{\mathbb{P}(Y_i=1|\X_i)}{\mathbb{P}(Y_i=0|\X_i)}=\beta_{0}+A_{\bmbeta}\X_i\bmbeta ~~\text{for}~ i \in [n].
\end{eqnarray*}
where $\bmbeta=(1_{s/3}\otimes(3,1.5,0,0,2,0,0),0_{p-7s/3})\odot \epsilon$ is a sparse vector with sparsity level $s=p/4$, $A_{\bmbeta}=0.5$ or $1.5$ controls the magnitude of the effect while $\epsilon$ is a vector of independent Rademacher variables, and $\beta_0$ controls the prevalence of outcome and is set as $\beta_0=-1$.

Third, we sample measurement error $\bmepsilon_w\sim \mathcal{N}(0,\mathbf{\Sigma}_{\epsilon})$ and calculate $\W=\X+\bmepsilon_w$. Here we consider $\mathbf{\Sigma}_{\epsilon}=AR(\sigma_{\epsilon},\rho_{\epsilon},p)$ where $\sigma^2_{\epsilon}=0$, $0.1$, $0.6$ or $1$ controls the scale of measurement errors, and $\rho_{\epsilon}$ controls the correlation between measurement errors, which is set at $\rho_{\epsilon}=0.3$. 

Fourth, we sample missing data indicators $R_{ij}$ under the missing at random (MAR) mechanism. We first randomly choose subsets $S_{mis0}\subset [p]\cap \mathcal{H}$ and $S_{mis1} \subset [p]\cap \mathcal{H}^c$ so that $|S_{mis0}|=\pi_{mis} \cdot (p-s)$ and $|S_{mis1}|=\pi_{mis} \cdot s$ where $\pi_{mis}$ controls the proportion of variables that will contain missing values and is set at $2/15$ to approximate the proportion of variables with more than 5\% missing in our real data set or at 0 for setting without missing data. Then for each $j\in S_{mis}=S_{mis0}\cup S_{mis1}$, we sample
$R_{ij}$ independently from a sequence of logistic regression models
\begin{eqnarray*}
\log \frac{\mathbb{P}(R_{ij}=1|\W_{i,-j})}{\mathbb{P}(R_{ij}=0|\W_{i,-j})}=\eta_{0j}+ \W_{i,-j}\bm{\eta}_j ,
\end{eqnarray*}
where the intercept $\eta_{0j}$ is used to control the average proportion of missing at 5\% or 15\% and elements of $\bm{\eta}_j$ are independently sampled from $\text{Uniform}[-2,2]$. For $j\notin S_{mis}$, we let $R_{ij}=1$. Additional simulation when the missing depend on $\X$ rather than $\W$ are also considered where we will then sample $R_{ij}$ for each $j\in S_{mis}$ by 
\begin{eqnarray*}
\log \frac{\mathbb{P}(R_{ij}=1|\X_{i,-j})}{\mathbb{P}(R_{ij}=0|\X_{i,-j})}=\eta_{0j}+\X_{i,-j}\bm{\eta}_j,
\end{eqnarray*}
where the intercept $\eta_{0j}$ is used to control the average proportion of missing among those variables with missing value at $p_{mis}=5\%$ or 15\% and elements of $\bm{\eta}_j$ are independently sampled from $\text{Uniform}[-2,2]$. 

We first consider a setting with $\mathbf{\Sigma}_{\epsilon}=\mathbf{0}$, which is a setting with only missing data but not measurement errors. Under this setting, we compare the performance of the following methods: Lasso, Lasso Order, RF in terms of the FDR control and power, calculated from 100 replicates when using multiple imputations from chained equations with 5 imputed datasets each. We considered either including or excluding the outcome $\Y$ when performing the imputations and consider three different imputation methods (\textit{default} method using generalized linear models, classification and regression tree (\textit{cart}), and random forest (\textit{rf}). 

The second setting is with $\pi_{mis}=0$, which reflects a setting with only measurement errors, but not missingness in the data. Under this setting, we compare the performance of the following methods: Lasso, Lasso Order, RF, GDS, GMUS, and Corrected Lasso in terms of FDR control and power calculated from 100 replicates. 

The third setting is with $\pi_{mis}\neq 0$ and $\mathbf{\Sigma}_{\epsilon}\neq \mathbf{0}$, which reflects a setting with both missing data and measurement errors. Under this setting, we compare the performance of the following methods: Lasso, Lasso Order, RF, GDS, GMUS, and Corrected Lasso with respect to FDR control and power calculated from 100 replicates when using multiple imputations from chained equations with 5 imputed datasets each. 

\subsection{Web Appendix A.2: Additional Simulation Settings and Results}\label{APP:A.2}
Since the missing at random assumption based on error-prone variables ($\W$) can be strong for some applications, here we present the simulation study to see how the result will be sensitive to that assumption when the truth is that the missing probability depends on the error-free variables ($\X$). The results are summarized in Table \ref{tab:sim3n}.

\begin{table}[h]
\centering
\caption{Simulation results (FDP and Power) of different methods for Setting 3 (data with both measurement errors and missing data) when the missing probability depends on the error-free variables $\X$.}
\label{tab:sim3n}
\begin{tabular}{ccccccccc}
\hline
\multicolumn{2}{c}{FDP} & & & \\ \hline
$\sigma_{\epsilon}^2$  & Imp M & Imp Y· & Lasso & Lasso Order & RF & GDS & Corrected Lasso & GMUS \\ \hline·
0.1       & default       & yes      & 0.34 & 0.23 & 0.12 & {\color{black}0.36} & 0.04  & {\color{black}0.31}\\
0.1       & default       & no       & 0.32 & 0.22 & 0.14 & {\color{black}0.33} & 0.03  & {\color{black}0.28}\\
0.1       & cart          & yes      & 0.32 & 0.24 & 0.12 & {\color{black}0.33} & 0.04  & {\color{black}0.28}\\
0.1       & cart          & no       & 0.30 & 0.23 & 0.15 & {\color{black}0.32} & 0.03  & {\color{black}0.26} \\
0.1       & rf            & yes      & 0.34 & 0.24 & 0.15 & {\color{black}0.34} & 0.04  & {\color{black}0.29}\\
0.1       & rf            & no       & 0.32 & 0.25 & 0.15 & {\color{black}0.32} & 0.02  & {\color{black}0.29}\\
0.6  & default & yes & 0.57 & 0.26 & 0.10 & {\color{black}0.54} & 0.07  & {\color{black}0.52}\\
0.6  & default & no & 0.56 & 0.27 & 0.13 & {\color{black}0.50} & 0.07  & {\color{black}0.49}\\
0.6  & cart & yes & 0.56 & 0.32 & 0.13 & {\color{black}0.54} & 0.08  & {\color{black}0.50}\\
0.6  & cart & no & 0.56 & 0.30 & 0.12 & {\color{black}0.52} & 0.08  & {\color{black}0.50}\\
0.6  & rf & yes & 0.56 & 0.32 & 0.13 & {\color{black}0.54} & 0.07  & {\color{black}0.49}\\
0.6  & rf & no & 0.56 & 0.30 & 0.13 & {\color{black}0.54} & 0.07  & {\color{black}0.50}\\ \hline
\multicolumn{2}{c}{Power} & & & \\ \hline
$\sigma_{\epsilon}^2$  & Imp M & Imp Y & Lasso & Lasso Order & RF & GDS & Corrected Lasso & GMUS \\ \hline
0.1  & default & yes & 1.00 & 0.93 & 0.76 & {\color{black}1.00} & 0.41 & {\color{black}1.00} \\
0.1  & default & no & 1.00 & 0.91 & 0.77 & {\color{black}1.00} & 0.36 & {\color{black}1.00}\\
0.1  & cart & yes & 1.00 & 0.92 & 0.78 & {\color{black}1.00} & 0.37 & {\color{black}1.00}\\
0.1  & cart & no & 1.00 & 0.91 & 0.78 & {\color{black}1.00}& 0.37 & {\color{black}1.00} \\
0.1  & rf & yes & 1.00 & 0.93 & 0.79 & {\color{black}1.00}& 0.37 & {\color{black}1.00}\\
0.1  & rf & no & 1.00 & 0.93 & 0.78 & {\color{black}1.00} & 0.33 & {\color{black}1.00}\\
0.6  & default & yes & 0.93 & 0.70 & 0.52 & {\color{black}0.93} & 0.32 & {\color{black}0.92} \\
0.6  & default & no & 0.92 & 0.73 & 0.56 & {\color{black}0.92} & 0.29 & {\color{black}0.92} \\
0.6  & cart & yes & 0.94 & 0.73 & 0.59 & {\color{black}0.94}& 0.31 & {\color{black}0.93} \\
0.6  & cart & no & 0.94 & 0.74 & 0.56 & {\color{black}0.94} & 0.29 & {\color{black}0.93} \\
0.6  & rf & yes & 0.94 & 0.77 & 0.60 & {\color{black}0.94} & 0.30 & {\color{black}0.93} \\
0.6  & rf & no & 0.94 & 0.75 & 0.58 & {\color{black}0.94}& 0.29 & {\color{black}0.93}\\
\hline
\end{tabular}
\end{table}

{\color{black}\subsection{Web Appendix A.3: Additional Simulation Settings and Results for General Simultaneous Knockoff Methods}\label{APP:A.3}
We generate two independent datasets with the same settings outlined in Table 3, except for the values for the beta vectors. First, we sample two feature matrices $X^1, X^2$ independently with the same as the description in Appendix A.1. Second, for beta, the mutual signals for both datasets are the same, and two non-mutual signals have magnitudes 0.5 and 1 with random direction in both data sets. 
\begin{eqnarray*}
\beta^1&=\beta(\mathbf{1}_{s/3}\otimes(3,1.5,0,0,2,0,0),\omega_1,0_{2},0_{p-7s/3-4})\odot \epsilon,\\
\beta^2&=\beta(\mathbf{1}_{s/3}\otimes(3,1.5,0,0,2,0,0),0_{2},\omega_2,0_{p-7s/3-4})\odot \epsilon,
\end{eqnarray*}
where $\omega_1=(0.5,1) \odot \epsilon_1$, $\omega_2=(0.5,1) \odot \epsilon_2$, $\epsilon_1,\epsilon_2,$ and $\epsilon$ are vectors of independent Rademacher variables.
Then $\Y^1, \Y^2$s are generated from logistic regression models
\begin{eqnarray*}
\log \frac{\mathbb{P}(Y_i^1=1|\X_i^1)}{\mathbb{P}(Y_i^1=0|\X_i^1)}=\beta_{0}+A_{\bmbeta}\X_i^1\bmbeta^1,
\log \frac{\mathbb{P}(Y_i^2=1|\X_i^2)}{\mathbb{P}(Y_i^2=0|\X_i^2)}=\beta_{0}+A_{\bmbeta}\X_i^2\bmbeta^2
~~\text{for}~ i \in [n].
\end{eqnarray*}
where $A_{\bmbeta}=1$ controls the magnitude of the effect, $\beta_0$ controls the prevalence of outcome and is set as $\beta_0=-1$. The measurement error and missing data indicators are independently sampled for both datasets and the same as the description in Appendix A.1.

We simulate both datasets using Setting 3 (both data with both missing data and measurement errors). The results are summarized in Table \ref{tab:simul_sim}. The same as Table 3, we include the variable selection methods Lasso, Lasso Order, RF, GDS, Corrected Lasso, and GMUS. We present the results where the missing probabilities depend on W rather than X here. We consider two levels of Scales for the measurement errors. For imputation methods, we consider
the default, cart, and rf. We also compare the performance based on whether to impute Y. 
When the scale of measurement errors is small (Scale = 0.1), all methods control the FDR under the nominal values of 0.2. With bigger scale measurement errors (Scale = 0.6), Lasso Order and RF methods also fail to control the FDR. Corrected Lasso consistently has the lowest FDR, followed by GMUS across all the designed settings. Lasso and GDS also control FDR across all the settings but the FDR are larger than Corrected Lasso and GMUS. In terms of power, the GDS and Lasso methods have the best performance, followed by GMUS and Lasso Order. Corrected Lasso does not have good power. The results for general simultaneous knockoff methods are consistent with the single dataset settings.

\begin{table}[h]
\centering
\caption{Simulation results (FDP and Power) for Simultaneous Knockoff Methods for Two datasets with both measurement errors and missing data) for $n=1000$, $p=60$, $p_{mis}=0.15$, and $A_{\bmbeta}=1$, varying $\sigma_{\epsilon}^2$, Imp M and Imp Y.}
\label{tab:simul_sim}
\begin{tabular}{ccccccccc}
\hline
\multicolumn{2}{c}{FDP} & & & \\ \hline
$\sigma_{\epsilon}^2$ & Imp M & Imp Y & Lasso & Lasso Order & RF & GDS & Corrected Lasso & GMUS \\ \hline
0.1       & default       & yes      & 0.09 & 0.18 & 0.18 & 0.11 & 0.00 & 0.05 \\
0.1       & default       & no       & 0.08 & 0.19 & 0.17 & 0.10 & 0.00 & 0.05 \\
0.1       & cart          & yes      & 0.08 & 0.18 & 0.17 & 0.11 & 0.00 & 0.05 \\
0.1       & cart          & no       & 0.08 & 0.19 & 0.18 & 0.10 & 0.00 & 0.04 \\
0.1       & rf            & yes      & 0.10 & 0.19 & 0.18 & 0.12 & 0.00 & 0.05 \\
0.1       & rf            & no       & 0.10 & 0.18 & 0.18 & 0.12 & 0.00 & 0.05 \\
0.6  & default & yes                 & 0.11 & 0.21 & 0.21 & 0.12 & 0.07 & 0.07 \\
0.6  & default & no                  & 0.11 & 0.22 & 0.23 & 0.12 & 0.08 & 0.07 \\
0.6  & cart & yes                    & 0.11 & 0.22 & 0.22 & 0.12 & 0.09 & 0.09 \\
0.6  & cart & no                     & 0.10 & 0.21 & 0.22 & 0.14 & 0.07 & 0.08 \\
0.6  & rf & yes                      & 0.11 & 0.23 & 0.21 & 0.13 & 0.08 & 0.09 \\
0.6  & rf & no                       & 0.12 & 0.24 & 0.23 & 0.14 & 0.08 & 0.08 \\
\hline
\multicolumn{2}{c}{Power} & & & \\ \hline
$\sigma_{\epsilon}^2$ & Imp M & Imp Y & Lasso & Lasso Order & RF & GDS & Corrected Lasso & GMUS \\ \hline
0.1  & default & yes &   1.00 & 0.96 & 0.85 & 1.00 & 0.22 & 0.98 \\
0.1  & default & no &    1.00 & 0.96 & 0.85 & 1.00 & 0.19 & 0.96 \\
0.1  & cart & yes &      1.00 & 0.95 & 0.85 & 1.00 & 0.19 & 0.96 \\
0.1  & cart & no &       1.00 & 0.96 & 0.84 & 1.00 & 0.20 & 0.95 \\
0.1  & rf & yes &        1.00 & 0.95 & 0.85 & 1.00 & 0.22 & 0.96 \\
0.1  & rf & no &         1.00 & 0.95 & 0.86 & 1.00 & 0.21 & 0.96 \\
0.6  & default & yes &   0.94 & 0.87 & 0.82 & 0.94 & 0.42 & 0.89 \\
0.6  & default & no &    0.93 & 0.87 & 0.81 & 0.93 & 0.42 & 0.89 \\
0.6  & cart & yes &      0.93 & 0.87 & 0.81 & 0.93 & 0.42 & 0.89 \\
0.6  & cart & no &       0.93 & 0.87 & 0.82 & 0.93 & 0.39 & 0.89 \\
0.6  & rf & yes &        0.93 & 0.87 & 0.82 & 0.93 & 0.41 & 0.89 \\
0.6  & rf & no &         0.93 & 0.87 & 0.82 & 0.93 & 0.42 & 0.89 \\
\hline
\end{tabular}

\end{table}
}
\section{Web Appendix B: Additional data analysis details}\label{APP:B}
\subsection{Web Appendix B.1: Matching Method}\label{APP:B.1}
Cases and controls for this analysis were selected from the entire Women's Health Initiative (WHI) Bone Mineral Density (BMD) Subcohort (n=11,020). The BMD was comprised of women in both the Clinical Trial (CT) and Observational Study (OS), who were enrolled at three specified WHI clinical centers (Birmingham, Pittsburgh, and Tucson/Phoenix), had dual X-ray absorptiometry at baseline and follow-up time points, and provided spot urine specimens, as well as fasting blood samples. {\color{black}For the CT samples, they include both dietary modification trials (DM) and hormone therapy trials (HT).} Here the eligible sample was restricted to women who had sufficient WHI serum (300 $\mu$l) and urine (550 $\mu$l) samples from the same time point, before and closest to the case diagnosis date and were required to have no missing covariate data (n=10,451). 
The cases were defined as the earliest incident invasive breast cancer (BC) or colorectal cancer (CRC) so that the biospecimen collection would be reasonably proximate. Each of the 758 case women was matched 1-to-1 to a control woman, disease-free at the case occurrence follow-up time, based on age (within 2 years; Table I), WHI enrollment date (within 2 months to control for follow-up duration), and race/ethnicity.  Participants could only be a control for one case, and a case could not be a control for another case. The matching algorithm was applied to select the closest match based on criteria to minimize an overall distance measure \citep{bergstrahl}. Each matching factor was given the same weight. Controls were excluded for the following reasons: a) history of BC or CRC reported at baseline (n=382); b) no follow-up (n=32); c) missing any covariate data (n=3513 breast, n=2905 colorectal). The number of eligible controls was n=6477 BC controls, n=7056 CRC controls. 

Because these two control groups overlap; the 181 CRC cases were matched first, matched controls were removed from the eligible pool, then BC cases were matched. 54\% of the selected sample were in the OS, 34\% in the DM, and 12\% in the HT-not DM.

\subsection{Web Appendix B.2: Data preprocessing}\label{APP:B.2}
Data are analyzed separately for each metabolomics platform. For metabolic variables, we removed those with more than 
20\% missing values. We take log transformation to all lab-measured variables to be consistent with other analyses in the Nutrition and Physical Activity Assessment Study Feeding Study (NPAAS-FS) \cite{zheng2021evaluation}. Outliers were truncated to Q1-3*IQR or Q3+3*IQR where Q1 and Q3 are the first and the third quartiles and IQR is the interquartile range. To remove the batch and run order effect for LC-MS and GC-MS data, normalization was performed using local polynomial regression fitting over run order within each batch. 

\subsection{Web Appendix B.3: Calculation of variance-covariance matrix for the measurement errors}\label{APP:B.3}
To calculate the variance-covariance matrix for the measurement errors, we utilize the QC samples. Specifically, we use pooled NPAAS-FS first void urine QC samples for the GC-MS platform and NMR platform and we use pooled NPAAS and NPAAS-FS serum QC samples for the LC-MS platform and the lipidomic platform. Log transformation is also performed on all the lab-measured variables to be consistent with non-QC samples. We also remove variables with more than 20\% missing values in non-QC samples and all missings in QC samples. The same normalization method is also performed for QC samples. For GC-MS and NMR, we use the sample variance and covariance matrix of the QC samples to estimate the variance-covariance matrix for the measurement errors. For lipidomic and NMR platforms, QC samples are collected twice for each batch. To fully remove the batch effect, we use half of the sample variance-covariance matrix of the difference within each batch to estimate the variance-covariance matrix for the measurement error. If the estimated variance-covariance matrix as above is not positive definite, we add a data-adaptive value (i.e., the first eigenvalue less than a threshold $10^{-4}$) to the corresponding correlation matrix and re-calculate the covariance matrix based on the new correlation matrix and original standard deviations. {\color{black}The summary information of missing and measurement error for each platform (after removing those with $>$ 20\% missing) is shown in Table \ref{tab:summary-table}.}

\begin{table}[h]
\centering
\caption{Summary information of missing and measurement errors for each platform}
\label{tab:summary-table}
\begin{adjustbox}{max width=\textwidth,max totalheight=\textheight,keepaspectratio}
\begin{tabular}{llll}
\hline
  \begin{tabular}[c]{@{}l@{}}Platform\end{tabular} &
  \begin{tabular}[c]{@{}l@{}}Number of variables\\ having missing\\N   (\%)\end{tabular} &
  \begin{tabular}[c]{@{}l@{}}The proportion (\%)\\ of missing\\Mean (SD)\end{tabular} &
  \begin{tabular}[c]{@{}l@{}}SNR* \\ Median, [Q1, Q3]\\ (Min$-$Max)\end{tabular} 
  \\\hline
  \begin{tabular}[c]{@{}l@{}} NMR \end{tabular}&
  \begin{tabular}[c]{@{}l@{}}10   (16.95) \end{tabular}&
  \begin{tabular}[c]{@{}l@{}}2.36   (2.67) \end{tabular}&
  \begin{tabular}[c]{@{}l@{}}100.7,   [58.2, 169.1]\\(8.0$-$1873.3) \end{tabular}\\\hline
  \begin{tabular}[c]{@{}l@{}}GCMS \end{tabular}&
  \begin{tabular}[c]{@{}l@{}}58 (86.57) \end{tabular}&
  \begin{tabular}[c]{@{}l@{}}6.25 (6.05) \end{tabular}&
  \begin{tabular}[c]{@{}l@{}} 34.0, [14.8, 92.1] \\(0.1$-$1397328.9) \end{tabular}\\\hline
  \begin{tabular}[c]{@{}l@{}}LCMS\\(AbsQuant) \end{tabular}&
  \begin{tabular}[c]{@{}l@{}}23 (76.67) \end{tabular}&
  \begin{tabular}[c]{@{}l@{}}0.07   (0) \end{tabular}&
  \begin{tabular}[c]{@{}l@{}}37.3, [13.8, 79.0]\\(1.0$-$1786.6) \end{tabular}\\\hline
  \begin{tabular}[c]{@{}l@{}}LCMS\\(RelQuant) \end{tabular}&
  \begin{tabular}[c]{@{}l@{}}91   (61.69) \end{tabular}&
  \begin{tabular}[c]{@{}l@{}}1.36   (3.6) \end{tabular}&
  \begin{tabular}[c]{@{}l@{}}92.4, [31.3, 251.8] \\(0.2$-$44362.3) \end{tabular} \\\hline
  \begin{tabular}[c]{@{}l@{}}Lipidyzer\\(composition)\end{tabular} &
  \begin{tabular}[c]{@{}l@{}}413   (60.82)\end{tabular} &
  \begin{tabular}[c]{@{}l@{}}2.14   (3.61) \end{tabular}&
  \begin{tabular}[c]{@{}l@{}}37.5, [14.8, 100.0] \\(0.9$-$2133.4) \end{tabular} \\\hline
  \begin{tabular}[c]{@{}l@{}}Lipidyzer\\(concentration) \end{tabular}&
  \begin{tabular}[c]{@{}l@{}}413   (60.82) \end{tabular}&
  \begin{tabular}[c]{@{}l@{}}2.14   (3.61) \end{tabular}&
  \begin{tabular}[c]{@{}l@{}}76.8, [28.3, 228.5]    \\(1.9$-$2695.9) \end{tabular}
  \\\hline
\end{tabular}
\end{adjustbox}\\
*SNR is signal noise ratio defined as $Var(X)/Var(\epsilon_W)$.
\end{table}

\section{Web Appendix C: Additional data analysis results}\label{APP:C}
In this section, we first provide the detailed list of metabolites selected for at least $10\%$ of times from the stability selection for each method and each cancer outcome separately. Then we provide the detailed list of metabolites selected for at least $10\%$ of times from the stability selection for associated with both breast and colorectal cancer using preferred methods (Lasso, Lasso order, and GMUS). Finally, we provide the selected for at least $50\%$ of times from the stability selection for the preferred method and each cancer outcome separately when using only matched controls specific to that cancer.

\subsection*{Data analysis results using Lasso}
Metabolites that are selected using Lasso among $\geq$10\% of replications associated with BC and CRC are listed in Tables \ref{tab:bc-lasso} and \ref{tab:crc-lasso}.
\begin{table}[h]
\centering
\caption{Metabolites that are selected using \textbf{Lasso} among $\geq$10\% of replications associated with BC risks and the direction of their marginal association to the BC risks.}
\label{tab:bc-lasso}
\begin{adjustbox}{max width=\textwidth,max totalheight=\textheight,keepaspectratio}
\begin{tabular}{lll}
\hline
  \begin{tabular}[c]{@{}l@{}}Platform\end{tabular} &
  \begin{tabular}[c]{@{}l@{}}Half Min Imputation\end{tabular} &
 \begin{tabular}[c]{@{}l@{}} Multiple Imputation\end{tabular} \\ \hline
  \begin{tabular}[c]{@{}l@{}}NMR\end{tabular} &
  \begin{tabular}[c]{@{}l@{}} \end{tabular} &
  \begin{tabular}[c]{@{}l@{}} \end{tabular} \\ \hline
  
  \begin{tabular}[c]{@{}l@{}}GCMS\end{tabular} &
  \begin{tabular}[c]{@{}l@{}}Alpha$-$ketoglutarate ({\color{black}76}\%)(-)\end{tabular} &
  \begin{tabular}[c]{@{}l@{}} \end{tabular} \\ \hline

  \begin{tabular}[c]{@{}l@{}}LCMS \\ (AbsQuant)\end{tabular} &
  \begin{tabular}[c]{@{}l@{}}Choline (13\%)(+)\\ 3HBA (13\%)(+)\end{tabular} &
  \begin{tabular}[c]{@{}l@{}} Choline (13\%)(+) \end{tabular} \\ \hline

  \begin{tabular}[c]{@{}l@{}}LCMS \\ (RelQuant)\end{tabular} &
  \begin{tabular}[c]{@{}l@{}} \end{tabular} &
  \begin{tabular}[c]{@{}l@{}} \end{tabular} \\\hline

  \begin{tabular}[c]{@{}l@{}}Lipidyzer \\ (composition)\end{tabular} &
  \begin{tabular}[c]{@{}l@{}}DAG 14:1/18:1 (40\%)(+)\\ TAG 47:0(FA15:0)) (39\%)(-)\\ TAG 48:3 (FA18:1) (24\%)(-)\end{tabular} &
  \begin{tabular}[c]{@{}l@{}}TAG 48:5(FA18:3) (80\%)(-)\\DAG 14:1/18:1 (78\%)(+)\\  TAG 48:0(FA16:0 )(20\%)(+)\end{tabular} \\\hline
   
  \begin{tabular}[c]{@{}l@{}}Lipidyzer \\ (concentration)\end{tabular}&
  \begin{tabular}[c]{@{}l@{}}DAG 14:1/18:1 (97\%)(+)\\ TAG 48:0(FA16:0) (49\%)(+)\\ TAG 56:9(FA20:4) (17\%)(-)\\ PE 18:1/20:3 (14\%)(-)\end{tabular}&
  \begin{tabular}[c]{@{}l@{}}DAG 14:1/18:1 (97\%)(+)\\ TAG48:5(FA18.3) (64\%)(-)\\ PE 18:2/20:4 (11\%)(-)\end{tabular} \\\hline
\end{tabular}
\end{adjustbox}
\end{table}

\begin{table}[h]
\centering
\caption{Metabolites that are selected using \textbf{Lasso} among $\geq$10\% of replications associated with CRC risks and the direction of their marginal association to the CRC risks.}
\label{tab:crc-lasso}
\begin{adjustbox}{max width=\textwidth,max totalheight=\textheight,keepaspectratio}
\begin{tabular}{lll}
\hline
  \begin{tabular}[c]{@{}l@{}} Platform \end{tabular} &
  \begin{tabular}[c]{@{}l@{}} Half Min Imputation \end{tabular} &
  \begin{tabular}[c]{@{}l@{}} Multiple Imputation \end{tabular} \\ \hline
  \begin{tabular}[c]{@{}l@{}}NMR \end{tabular}&
  \begin{tabular}[c]{@{}l@{}}N$-$methylnicotinic acid (43\%)(-)\\ Taurine (22\%)(+)\end{tabular} &
  \begin{tabular}[c]{@{}l@{}}Taurine (16\%)(+)\end{tabular} \\ \hline
  
  \begin{tabular}[c]{@{}l@{}}GCMS\end{tabular} &
  \begin{tabular}[c]{@{}l@{}}2,3$-$Dihydroxybutanoic acid ({\color{black}46}\%)(+)\end{tabular} &
  \begin{tabular}[c]{@{}l@{}}2,3$-$Dihydroxybutanoic acid ({\color{black}93}\%)(+)\end{tabular} \\ \hline
  
  \begin{tabular}[c]{@{}l@{}}LCMS \\ (AbsQuant)\end{tabular} &
  \begin{tabular}[c]{@{}l@{}}Glucose (84\%)(+)\\ Cystine (52\%)(-)\\ Serine (44\%)(+)\\ Urate (42\%)(+)\\ Choline (32\%)(+)\\ Glycine (18\%)(+)\\ Proline (11\%)(+)\end{tabular} &
  \begin{tabular}[c]{@{}l@{}}Glucose (62\%)(+)\\ Serine (43\%)(+)\\Urate (14\%)(+)\\ Choline (10\%)(+)\end{tabular} \\ \hline

  \begin{tabular}[c]{@{}l@{}}LCMS \\ (RelQuant)\end{tabular} &
  \begin{tabular}[c]{@{}l@{}}Glycerate (69\%)(+)\\ Adenosine (66\%)(-)\\ Adipic Acid (14\%)(+)\end{tabular}
&
  \begin{tabular}[c]{@{}l@{}}Adenosine (69\%)(-)\\Glycerate (67\%)(+)\end{tabular} \\\hline

  \begin{tabular}[c]{@{}l@{}}Lipidyzer \\ (composition)\end{tabular} &
  \begin{tabular}[c]{@{}l@{}}TAG 48:5(FA18:2)  (59\%)(+)\\ TAG 47:2(FA14:0)  (36\%)(-)\\ TAG 52:8(FA16:1)  (32\%)(-)\\ TAG 54:0(FA16:0)  (25\%)(-)\\ TAG 46:4(FA18:2)  (17\%)(+)\end{tabular} &
  \begin{tabular}[c]{@{}l@{}} \end{tabular} \\ \hline
   
  \begin{tabular}[c]{@{}l@{}}Lipidyzer \\ (concentration)\end{tabular} &
  \begin{tabular}[c]{@{}l@{}} \end{tabular}&
  \begin{tabular}[c]{@{}l@{}}  \end{tabular} \\ \hline
\end{tabular}

\end{adjustbox}
\end{table}

\subsection*{Data analysis results using Lasso Order}
Metabolites that are selected using Lasso Order among $\geq$10\% of replications associated with BC and CRC are listed in Tables \ref{tab:bc-lo} and \ref{tab:crc-lo}.
\begin{table}[h]
\centering
\caption{Metabolites that are selected using \textbf{Lasso Order} among $\geq$10\% of replications associated with BC risks and the direction of their marginal association to the BC risks.}
\label{tab:bc-lo}
\begin{adjustbox}{max width=\textwidth,max totalheight=\textheight,keepaspectratio}
\begin{tabular}{lll}
\hline
  \begin{tabular}[c]{@{}l@{}}Platform\end{tabular} &
  \begin{tabular}[c]{@{}l@{}}Half Min Imputation\end{tabular} &
  \begin{tabular}[c]{@{}l@{}}Multiple Imputation\end{tabular} \\ \hline
  \begin{tabular}[c]{@{}l@{}}NMR\end{tabular} &
  \begin{tabular}[c]{@{}l@{}}N$-$methylnicotinic acid (59\%)(+)\end{tabular} &
  \begin{tabular}[c]{@{}l@{}}N$-$methylnicotinic acid (63\%)(+)\end{tabular} \\ \hline

  \begin{tabular}[c]{@{}l@{}}GCMS\end{tabular} &
  \begin{tabular}[c]{@{}l@{}}Alpha$-$ketoglutarate ({\color{black}54}\%)(-)\end{tabular} &
  \begin{tabular}[c]{@{}l@{}}2,3$-$Dihydroxybutanoic acid ({\color{black}42}\%)(-)\end{tabular} \\ \hline

  \begin{tabular}[c]{@{}l@{}}LCMS \\ (AbsQuant)\end{tabular} &
  \begin{tabular}[c]{@{}l@{}}3HBA (67\%)(+)\\ Cystine (66\%)(-)\end{tabular} &
  \begin{tabular}[c]{@{}l@{}} Cystine (34\%)(-)\\ 3HBA (19\%)(+) \end{tabular} \\ \hline

  \begin{tabular}[c]{@{}l@{}}LCMS \\ (RelQuant)\end{tabular} &
  \begin{tabular}[c]{@{}l@{}}Malate (45\%) (-)\end{tabular}
&
  \begin{tabular}[c]{@{}l@{}}Malate (47\%)(-) \end{tabular} \\\hline

  \begin{tabular}[c]{@{}l@{}}Lipidyzer \\ (composition)\end{tabular} &
  \begin{tabular}[c]{@{}l@{}}TAG 47:0(FA15:0) (68\%)(-) \end{tabular} &
  \begin{tabular}[c]{@{}l@{}}TAG 48:5(FA18:3) (80\%)(-)\\ DAG 14:1/18:1 (76\%)(+) \\TAG 58:10(FA20:5) (37\%)(-)\\ TAG 56:9(FA20:4) (21\%)(-)\\ TAG 46:3(FA16:1) (16\%)(-)\end{tabular} \\ \hline
   
  \begin{tabular}[c]{@{}l@{}}Lipidyzer \\ (concentration)\end{tabular} &
  \begin{tabular}[c]{@{}l@{}}TAG 44:1(FA12:0) (24\%)(-)\end{tabular}&
  \begin{tabular}[c]{@{}l@{}} \end{tabular} \\ \hline
\end{tabular}
\end{adjustbox}
\end{table}

\begin{table}[h]
\centering
\caption{Metabolites that are selected using \textbf{Lasso Order} among $\geq$10\% of replications associated with CRC risks and the direction of their marginal association to the CRC risks.}
\label{tab:crc-lo}
\begin{adjustbox}{max width=\textwidth,max totalheight=\textheight,keepaspectratio}
\begin{tabular}{lll}
\hline
 \begin{tabular}[c]{@{}l@{}} Platform\end{tabular} &
  \begin{tabular}[c]{@{}l@{}}Half Min Imputation \end{tabular}&
  \begin{tabular}[c]{@{}l@{}}Multiple Imputation \end{tabular}\\\hline
  \begin{tabular}[c]{@{}l@{}} NMR \end{tabular}&
  \begin{tabular}[c]{@{}l@{}}N$-$methylnicotinic acid (76\%)(-) \end{tabular}&
  \begin{tabular}[c]{@{}l@{}}N$-$methylnicotinic acid (54\%)(-) \end{tabular}\\ \hline
 
  \begin{tabular}[c]{@{}l@{}} GCMS \end{tabular}&
  \begin{tabular}[c]{@{}l@{}}2,3$-$Dihydroxybutanoic acid ({\color{black}46}\%)(+)\end{tabular} &
  \begin{tabular}[c]{@{}l@{}}2,3$-$Dihydroxybutanoic acid ({\color{black}63}\%)(+)\end{tabular} \\ \hline
 
  \begin{tabular}[c]{@{}l@{}}LCMS \\ (AbsQuant)\end{tabular} &
  \begin{tabular}[c]{@{}l@{}}Cystine (100\%)(-)\\ 3HBA (75\%)(+)\\ Glutamic acid (18\%)(+)\\ Glucose (16\%)(+)\\ Glycine (12\%)(+)\\ Choline (12\%)(+)\\ Urate (12\%)(+)\end{tabular} &
  \begin{tabular}[c]{@{}l@{}}Cystine (94\%)(-)\\ 3HBA (31\%)(+)\end{tabular} \\ \hline
 
  \begin{tabular}[c]{@{}l@{}}LCMS \\ (RelQuant)\end{tabular} &
  \begin{tabular}[c]{@{}l@{}}Malate (57\%)(+) \end{tabular}&
  \begin{tabular}[c]{@{}l@{}}Malate (49\%)(+) \end{tabular}\\ \hline
 
  \begin{tabular}[c]{@{}l@{}}Lipidyzer \\ (composition)\end{tabular} &
  \begin{tabular}[c]{@{}l@{}}TAG 48:5(FA18:2)  (60\%)(+)\\ TAG 54:0(FA16:0)  (40\%)(-)\\ TAG 52:8(FA16:1)  (34\%)(-)\\ TAG 47:2(FA14:0)  (20\%)(-)\\ TAG 46:4(FA18:2)  (13\%)(+)\end{tabular} &
  \begin{tabular}[c]{@{}l@{}}TAG 48:4(FA14:0) (21\%)(-)\end{tabular} \\ \hline
 
  \begin{tabular}[c]{@{}l@{}}Lipidyzer \\ (concentration)\end{tabular} &
  \begin{tabular}[c]{@{}l@{}}TAG 47:2(FA14:0) (61\%)(-)\end{tabular} &
  \begin{tabular}[c]{@{}l@{}} \end{tabular} \\ \hline
\end{tabular}
\end{adjustbox}
\end{table}
\subsection*{Data analysis results using Random Forest}
Metabolites that are selected using RF among $\geq$10\% of replications associated with BC and CRC are listed in Tables \ref{tab:bc-rf} and \ref{tab:crc-rf}.
\begin{table}[h]
\centering
\caption{Metabolites that are selected using \textbf{Random Forest} among $\geq$10\% of replications associated with BC risks and the direction of their marginal association to the BC risks.}
\label{tab:bc-rf}
\begin{adjustbox}{max width=\textwidth,max totalheight=\textheight,keepaspectratio}
\begin{tabular}{lll}
\hline
  \begin{tabular}[c]{@{}l@{}}Platform\end{tabular} &
  \begin{tabular}[c]{@{}l@{}}Half Min Imputation\end{tabular} &
  \begin{tabular}[c]{@{}l@{}}Multiple Imputation\end{tabular} \\ \hline
  \begin{tabular}[c]{@{}l@{}}NMR\end{tabular} &
  \begin{tabular}[c]{@{}l@{}}Uracil  (72\%)(-)\end{tabular} &
  \begin{tabular}[c]{@{}l@{}}Uracil (35\%)(-)\end{tabular} \\ \hline

  \begin{tabular}[c]{@{}l@{}}GCMS\end{tabular} &
  \begin{tabular}[c]{@{}l@{}} \end{tabular} &
  \begin{tabular}[c]{@{}l@{}} \end{tabular} \\ \hline

  \begin{tabular}[c]{@{}l@{}}LCMS \\ (AbsQuant)\end{tabular} &
  \begin{tabular}[c]{@{}l@{}}Choline (38\%)(+)\\ Citrulline (32\%)(-)\\Cystine (18\%)(-)\\
  3HBA (11\%)(+)\\\\\end{tabular} &
  \begin{tabular}[c]{@{}l@{}} Choline (12\%)(+)\end{tabular} \\ \hline

  \begin{tabular}[c]{@{}l@{}}LCMS \\ (RelQuant)\end{tabular} &
  \begin{tabular}[c]{@{}l@{}}  \end{tabular}
&
  \begin{tabular}[c]{@{}l@{}}Glycochenodeoxycholate (33\%)(+)\end{tabular} \\\hline

  \begin{tabular}[c]{@{}l@{}}Lipidyzer \\ (composition)\end{tabular} &
  \begin{tabular}[c]{@{}l@{}}DAG 14:1/18:1 (44\%)(+)\\PE 18:0/20:2 (25\%)(+)\end{tabular} &
  \begin{tabular}[c]{@{}l@{}}DAG 14:1/18:1 (23\%)(+)\end{tabular} \\ \hline
   
  \begin{tabular}[c]{@{}l@{}}Lipidyzer \\ (concentration)\end{tabular} &
  \begin{tabular}[c]{@{}l@{}}DAG 14:1/18:1 (67\%)(+)\\PEP 18:1/22:5 (16\%)(-)\end{tabular}&
  \begin{tabular}[c]{@{}l@{}}DAG 14:1/18:1 (82\%)(+)\\PE 18:2/20:4 (13\%)(-)\\PEO 18:0/18:1 (11\%)(-)\\PC 18:0/20:0 (10\%)(-)\\\end{tabular} \\ \hline
\end{tabular}
\end{adjustbox}
\end{table}

\begin{table}[h]
\centering
\caption{Metabolites that are selected using \textbf{Random Forest} among $\geq$10\% of replications associated with CRC risks and the direction of their marginal association to the CRC risks.}
\label{tab:crc-rf}
\begin{adjustbox}{max width=\textwidth,max totalheight=\textheight,keepaspectratio}
\begin{tabular}{lll}
\hline
 \begin{tabular}[c]{@{}l@{}} Platform\end{tabular} &
  \begin{tabular}[c]{@{}l@{}}Half Min Imputation \end{tabular}&
  \begin{tabular}[c]{@{}l@{}}Multiple Imputation \end{tabular}\\ \hline
  \begin{tabular}[c]{@{}l@{}} NMR \end{tabular}&
  \begin{tabular}[c]{@{}l@{}} \end{tabular}&
  \begin{tabular}[c]{@{}l@{}}  \end{tabular}\\ \hline
 
  \begin{tabular}[c]{@{}l@{}} GCMS \end{tabular}&
  \begin{tabular}[c]{@{}l@{}} \end{tabular} &
  \begin{tabular}[c]{@{}l@{}} \end{tabular} \\ \hline
 
  \begin{tabular}[c]{@{}l@{}}LCMS \\ (AbsQuant)\end{tabular} &
  \begin{tabular}[c]{@{}l@{}}Glucose (73\%)(+)\\Cystine (55\%)(-)\\Pentothenate (54\%)(+)\\Histidine (43\%)(-)\\Threonine  (19\%)(-)\\Serine (11\%)(+)\end{tabular} &
  \begin{tabular}[c]{@{}l@{}}Glucose (73\%)(+)\\Histidine (44\%)(-)\\Cystine (44\%)(-)\\Pentothenate (35\%)(+)\\Threonine (20\%)(-)\\\end{tabular} \\ \hline
 
  \begin{tabular}[c]{@{}l@{}}LCMS \\ (RelQuant)\end{tabular} &
  \begin{tabular}[c]{@{}l@{}}  \end{tabular}&
  \begin{tabular}[c]{@{}l@{}}Adenosine (24\%)(-) \end{tabular}\\ \hline
 
  \begin{tabular}[c]{@{}l@{}}Lipidyzer \\ (composition)\end{tabular} &
  \begin{tabular}[c]{@{}l@{}}LCER 16:0 (14\%)(-)\\PC 18:0/18:0 (13\%)(+)\\TAG 50:5(FA18:1) (10\%)(-)\end{tabular} &
  \begin{tabular}[c]{@{}l@{}}PC 18:0/18:0 (28\%)(+)\\LCER 16:0 (23\%)(-)\end{tabular} \\ \hline
 
  \begin{tabular}[c]{@{}l@{}}Lipidyzer \\ (concentration)\end{tabular} &
  \begin{tabular}[c]{@{}l@{}}PC 18:1/18:3 (21\%)(-)\\LPE 18:0 (14\%)(+)\\PC 18:2/18:3 (13\%)(-)\end{tabular} &
  \begin{tabular}[c]{@{}l@{}}PEO 16:0/18:2 (14\%)(-)\\LPE 20:4 (12\%)(+)\\PC 18:1/18:3 (11\%)(-)\end{tabular} \\ \hline
\end{tabular}
\end{adjustbox}
\end{table}

\subsection*{Data analysis results using GDS}
Metabolites that are selected using GDS among $\geq$10\% of replications associated with BC and CRC are listed in Tables \ref{tab:bc-lo} and \ref{tab:crc-lo}.
\begin{table}[h]
\centering
\caption{Metabolites that are selected using \textbf{GDS} among $\geq$10\% of replications associated with BC risks and the direction of their marginal association to the BC risks.}
\label{tab:bc-gds}
\begin{adjustbox}{max width=\textwidth,max totalheight=\textheight,keepaspectratio}
\begin{tabular}{lll}
\hline
  \begin{tabular}[c]{@{}l@{}} Platform \end{tabular} &
  \begin{tabular}[c]{@{}l@{}} Half Min Imputation \end{tabular} &
  \begin{tabular}[c]{@{}l@{}} Multiple Imputation \end{tabular} \\ \hline
  \begin{tabular}[c]{@{}l@{}}NMR\end{tabular} &
  \begin{tabular}[c]{@{}l@{}}Uracil (21\%)(-) \\ Formate (19\%)(+)\end{tabular} &
  \begin{tabular}[c]{@{}l@{}} \end{tabular} \\ \hline
  
  \begin{tabular}[c]{@{}l@{}}GCMS\end{tabular} &
  \begin{tabular}[c]{@{}l@{}} \end{tabular} &
  \begin{tabular}[c]{@{}l@{}} \end{tabular} \\ \hline

  \begin{tabular}[c]{@{}l@{}}LCMS \\ (AbsQuant)\end{tabular} &
  \begin{tabular}[c]{@{}l@{}}Choline (61\%)(+)  \end{tabular} &
  \begin{tabular}[c]{@{}l@{}}Choline (27\%)(+)  \end{tabular} \\ \hline

  \begin{tabular}[c]{@{}l@{}}LCMS \\ (RelQuant)\end{tabular} &
  \begin{tabular}[c]{@{}l@{}}  \end{tabular}
&
  \begin{tabular}[c]{@{}l@{}} \end{tabular} \\\hline

  \begin{tabular}[c]{@{}l@{}}Lipidyzer \\ (composition)\end{tabular} &
  \begin{tabular}[c]{@{}l@{}}TAG 52:2(FA18:2)  (68\%)(+)\\ PE 18:1/20:3 (36\%)(-)\\ TAG 50:4(FA18:1) (20\%)(-) \end{tabular} &
  \begin{tabular}[c]{@{}l@{}}PC 16:0/18:2 (52\%)(+)\\ TAG 52:2(FA18:2) (34\%)(+)\\ FFA 20:2 (13\%)(+)\end{tabular} \\ \hline
   
  \begin{tabular}[c]{@{}l@{}}Lipidyzer \\ (concentration)\end{tabular} &
  \begin{tabular}[c]{@{}l@{}}DAG 14:1/18:1 (64\%)(+)\\ PC 18:1/22:5 (40\%)(-)\\ SM 20:0 (25\%)(+)\\ FFA 20:2 (12\%)(+)\end{tabular}&
  \begin{tabular}[c]{@{}l@{}}DAG 14:1/18:1 (86\%)(+)\\ PE 18:2/20:4 (44\%)(-)\\ SM 20:0 (42\%)(+)\\ PC 18:1/22:5 (34\%)(-)\end{tabular} \\ \hline
\end{tabular}
\end{adjustbox}
\end{table}

\begin{table}[h]
\centering
\caption{Metabolites that are selected using \textbf{GDS} among $\geq$10\% of replications associated with CRC risks and the direction of their marginal association to the CRC risks.}
\label{tab:crc-gds}
\begin{adjustbox}{max width=\textwidth,max totalheight=\textheight,keepaspectratio}
\begin{tabular}{lll}
\hline
  \begin{tabular}[c]{@{}l@{}}Platform \end{tabular}&
  \begin{tabular}[c]{@{}l@{}}Half Min Imputation \end{tabular} &
  \begin{tabular}[c]{@{}l@{}}Multiple Imputation \end{tabular} \\ \hline
  \begin{tabular}[c]{@{}l@{}}NMR \end{tabular} &
  \begin{tabular}[c]{@{}l@{}}Taurine (51\%)(+)\\ Histidine (24\%)(-)\end{tabular} &
  \begin{tabular}[c]{@{}l@{}} Taurine (14\%)(+) \end{tabular} \\\hline
  
  \begin{tabular}[c]{@{}l@{}} GCMS \end{tabular} &
  \begin{tabular}[c]{@{}l@{}}   \end{tabular} &
  \begin{tabular}[c]{@{}l@{}} \end{tabular} \\ \hline
  
  \begin{tabular}[c]{@{}l@{}}LCMS \\ (AbsQuant)\end{tabular} &
  \begin{tabular}[c]{@{}l@{}}Serine (90\%)(+)\\ Histidine (89\%)(-)\\ Choline (60\%)(+)\\ Glucose (55\%)(+)\\ Urate (40\%)(+)\\  Glutamic acid (17\%)(+)\end{tabular} &
  \begin{tabular}[c]{@{}l@{}}Histidine (90\%)(-)\\Serine (74\%)(+)\\  Glucose (33\%)(+)\\ Choline (25\%)(+)\\ Urate (15\%)(+)\\ Glutamic acid (11\%)(+)\end{tabular} \\ \hline
  
  \begin{tabular}[c]{@{}l@{}}LCMS \\ (RelQuant)\end{tabular} &
  \begin{tabular}[c]{@{}l@{}}Adenosine (76\%)(-)\\ Glycerate (19\%)(+)\end{tabular} &
  \begin{tabular}[c]{@{}l@{}}Adenosine (84\%)(-)\end{tabular} \\ \hline
  
  \begin{tabular}[c]{@{}l@{}}Lipidyzer \\ (composition)\end{tabular} &
  \begin{tabular}[c]{@{}l@{}}HCER 24:0 (21\%)(-)\\ TAG 54:2(FA18:1) (18\%)(+)\end{tabular} &
  \begin{tabular}[c]{@{}l@{}} \end{tabular} \\ \hline
  
  \begin{tabular}[c]{@{}l@{}}Lipidyzer \\ (concentration)\end{tabular} &
  \begin{tabular}[c]{@{}l@{}}CER 16:0 (53\%)(+)\\ PC 18:2/20.3 (20\%)(-)\\ CER 24:1 (18\%)(+)\end{tabular} &
  \begin{tabular}[c]{@{}l@{}}CER 16:0 (70\%)(+)\\ PC 18:2/20:3 (11\%)(-)\end{tabular} \\ \hline
\end{tabular}
\end{adjustbox}
\end{table}

\subsection*{Data analysis results using GMUS}
Metabolites that are selected using GMUS among $\geq$10\% of replications associated with BC and CRC are listed in Tables \ref{tab:bc-gmus} and \ref{tab:crc-gmus}.

\begin{table}[h]
\centering
\caption{Metabolites that are selected using \textbf{GMUS} among $\geq$10\% of replications associated with BC risks and the direction of their marginal association to the BC risks.}
\label{tab:bc-gmus}
\begin{adjustbox}{max width=\textwidth,max totalheight=\textheight,keepaspectratio}
\begin{tabular}{lll}
\hline
  \begin{tabular}[c]{@{}l@{}}Platform\end{tabular} &
  \begin{tabular}[c]{@{}l@{}}Half Min Imputation\end{tabular} &
  \begin{tabular}[c]{@{}l@{}}Multiple Imputation\end{tabular} \\ \hline
  \begin{tabular}[c]{@{}l@{}}NMR\end{tabular} &
  \begin{tabular}[c]{@{}l@{}}Uracil (19\%)(-)\\ Formate (11\%)(+)\end{tabular} &
  \begin{tabular}[c]{@{}l@{}} \end{tabular} \\ \hline
  
  \begin{tabular}[c]{@{}l@{}}GCMS\end{tabular} &
  \begin{tabular}[c]{@{}l@{}} \end{tabular} &
  \begin{tabular}[c]{@{}l@{}} \end{tabular} \\ \hline

  \begin{tabular}[c]{@{}l@{}}LCMS \\ (AbsQuant)\end{tabular} &
  \begin{tabular}[c]{@{}l@{}}Choline (46\%)(+)\end{tabular} &
  \begin{tabular}[c]{@{}l@{}}Choline (38\%)(+)\end{tabular} \\ \hline

  \begin{tabular}[c]{@{}l@{}}LCMS \\ (RelQuant)\end{tabular} &
  \begin{tabular}[c]{@{}l@{}}  \end{tabular}
&
  \begin{tabular}[c]{@{}l@{}}  \end{tabular} \\\hline

  \begin{tabular}[c]{@{}l@{}}Lipidyzer \\ (composition)\end{tabular} &
  \begin{tabular}[c]{@{}l@{}}TAG 52:2(FA18:2)  (66\%)(-)\\ PE 18:1/20:3 (21\%)(-)\end{tabular} &
  \begin{tabular}[c]{@{}l@{}}PC 16:0/18:2 (52.6\%)(+)\\ TAG 52:2(FA18:2) (24.2\%)(+)\end{tabular} \\ \hline
   
  \begin{tabular}[c]{@{}l@{}}Lipidyzer \\ (concentration)\end{tabular} &
  \begin{tabular}[c]{@{}l@{}}DAG 14:1/18:1 (63\%)(+)\\ PC 18:1/22:5 (42\%)(-)\\ TAG 54:8(FA20:4)(13\%)(-)\\ FFA 20:2 (11\%)(+)\\ SM 20:0 (11\%)(+)\end{tabular}&
  \begin{tabular}[c]{@{}l@{}}DAG 14:1/18:1 (93.3\%)(+)\\ PE 18:2/20:4 (37.8\%)(-)\\ PC 18:1/22:5 (31.1\%)(-)\\ SM 20:0 (13.3\%)(+)\end{tabular} \\\hline
\end{tabular}
\end{adjustbox}
\end{table}

\begin{table}[h]
\centering
\caption{Metabolites that are selected using \textbf{GMUS} among $\geq$10\% of replications associated with CRC risks and the direction of their marginal association to the CRC risks.}
\label{tab:crc-gmus}
\begin{adjustbox}{max width=\textwidth,max totalheight=\textheight,keepaspectratio}
\begin{tabular}{lll}
\hline
  \begin{tabular}[c]{@{}l@{}} Platform \end{tabular} &
  \begin{tabular}[c]{@{}l@{}} Half Min Imputation \end{tabular} &
  \begin{tabular}[c]{@{}l@{}} Multiple Imputation \end{tabular} \\ \hline
  \begin{tabular}[c]{@{}l@{}} NMR  \end{tabular}&
  \begin{tabular}[c]{@{}l@{}}Taurine (35\%)(+)\\ Histidine (19\%)(-)\end{tabular} &
  \begin{tabular}[c]{@{}l@{}} Taurine (28\%)(+) \end{tabular}\\\hline
 
  \begin{tabular}[c]{@{}l@{}}  GCMS \end{tabular} &
  \begin{tabular}[c]{@{}l@{}}     \end{tabular} & 
  \begin{tabular}[c]{@{}l@{}} \end{tabular} \\\hline
  
  \begin{tabular}[c]{@{}l@{}}LCMS \\ (AbsQuant)\end{tabular} &
  \begin{tabular}[c]{@{}l@{}}Choline (42\%)(+)\\ Glucose (39\%)(+)\\ Serine (35\%)(+)\\ Histidine (35\%)(-)\\ Cystine (31\%)(-)\\ Glutamic acid (25\%)(+)\\ Urate (12\%)(+)\end{tabular} &
  \begin{tabular}[c]{@{}l@{}}Histidine (69\%)(-)\\Serine (46\%)(+)\\Glucose (25\%)(+)\\Choline (23\%)(+)\\ Glutamic acid (16\%)(+)\end{tabular} \\ \hline
  
  \begin{tabular}[c]{@{}l@{}}LCMS \\ (RelQuant)\end{tabular} &
  \begin{tabular}[c]{@{}l@{}}Adenosine (42\%)(-)\\ Glycerate (28\%)(+)\end{tabular} &
  \begin{tabular}[c]{@{}l@{}}Adenosine (29\%)(-)\\ Glycerate (11\%)(+)\end{tabular} \\\hline
   
  \begin{tabular}[c]{@{}l@{}}Lipidyzer \\ (composition)\end{tabular} &
  \begin{tabular}[c]{@{}l@{}}HCER 24:0 (12\%)(-)\\ TAG 54:2(FA18:1) (10\%)(+)\end{tabular} &
  \begin{tabular}[c]{@{}l@{}} \end{tabular} \\ \hline
  
  \begin{tabular}[c]{@{}l@{}}Lipidyzer \\ (concentration)\end{tabular} &
  \begin{tabular}[c]{@{}l@{}}CER 16:0 (52\%)(+)\\ PC 18:2/20:3 (22\%)(-)\\ CER 24:1 (15\%)(+)\end{tabular} &
  \begin{tabular}[c]{@{}l@{}}CER 16:0 (66\%)(+)\end{tabular} \\\hline
\end{tabular}
\end{adjustbox}
\end{table}

\subsection*{Data analysis results using Corrected Lasso}
Metabolites that are selected using Corrected Lasso among $\geq$10\% of replications associated with BC and CRC are listed in Tables \ref{tab:bc-cl} and \ref{tab:crc-cl}.

\begin{table}[h]
\centering
\caption{Metabolites that are selected using \textbf{Corrected Lasso} among $\geq$10\% of replications associated with BC risks and the direction of their marginal association to the BC risks.}
\label{tab:bc-cl}
\begin{adjustbox}{max width=\textwidth,max totalheight=\textheight,keepaspectratio}
\begin{tabular}{lll}
\hline
  \begin{tabular}[c]{@{}l@{}} Platform \end{tabular} &
  \begin{tabular}[c]{@{}l@{}} Half Min Imputation \end{tabular} &
  \begin{tabular}[c]{@{}l@{}} Multiple Imputation \end{tabular} \\ \hline
  \begin{tabular}[c]{@{}l@{}}NMR\end{tabular} &
  \begin{tabular}[c]{@{}l@{}}N$-$methylnicotinic acid (95\%)(+) \end{tabular} &
  \begin{tabular}[c]{@{}l@{}}N$-$methylnicotinic acid (99\%)(+) \end{tabular} \\ \hline
  
  \begin{tabular}[c]{@{}l@{}}GCMS\end{tabular} &
  \begin{tabular}[c]{@{}l@{}}Alpha$-$ketoglutarate ({\color{black}55}\%)(-)\\ 2,3$-$Dihydroxybutanoic acid ({\color{black}35}\%)(-)\\Serine ({\color{black}21}\%)(-)\\Phenol ({\color{black}15}\%)(-)\end{tabular} &
  \begin{tabular}[c]{@{}l@{}}Alpha$-$ketoglutarate ({\color{black}73}\%)(-)\\2,3$-$Dihydroxybutanoic acid ({\color{black}11}\%)(-)\end{tabular} \\ \hline

  \begin{tabular}[c]{@{}l@{}}LCMS \\ (AbsQuant)\end{tabular} &
  \begin{tabular}[c]{@{}l@{}}Aspartic Acid (17\%)(+)\\ Glucose (15\%)(+)\\ Cystine (11\%)(-)\\\end{tabular} &
  \begin{tabular}[c]{@{}l@{}} \end{tabular} \\ \hline

  \begin{tabular}[c]{@{}l@{}}LCMS \\ (RelQuant)\end{tabular} &
  \begin{tabular}[c]{@{}l@{}}Ribose$-$5$-$P (43\%)(-)\\ Malate (12\%)(-)\end{tabular}
&
  \begin{tabular}[c]{@{}l@{}}Ribose$-$5$-$P (30\%)(-)\end{tabular} \\ \hline

  \begin{tabular}[c]{@{}l@{}}Lipidyzer \\ (composition)\end{tabular} &
  \begin{tabular}[c]{@{}l@{}}DAG 14:1/18:1 (25\%)(+)\\ TAG 47:0(FA15:0) (18\%)(-)\\ PC 18:1/22:4 (13\%)(-)\\ PEP 18:1/22:4 (13\%)(+)\\ DAG 16:1/18:1 (12\%)(-)\\ DAG 18:0/18:1 (12\%)(+)\\ DAG 18:2/20:4 (10\%)(-)\\TAG 44:0(FA16:0) (10\%)(-)\\ TAG 44:1(FA12:0) (10\%)(-) \end{tabular} &
  \begin{tabular}[c]{@{}l@{}} \end{tabular} \\ \hline
   
  \begin{tabular}[c]{@{}l@{}}Lipidyzer \\ (concentration)\end{tabular} &
  \begin{tabular}[c]{@{}l@{}}TAG 44:0(FA16:0) (34\%)(-)\\ TAG 44:1(FA12:0) (15\%)(-)\\ TAG 44:0(FA14:0) (13\%)(-)\\ TAG 46:0(FA16:0) (12\%)(-)\end{tabular}&
  
  \begin{tabular}[c]{@{}l@{}}TAG 44:0(FA16:0) (32\%)(-)\\ TAG 46:1(FA16:1) (10\%) (-)\end{tabular} \\ \hline
\end{tabular}
\end{adjustbox}
\end{table}

\begin{table}[h]
\centering
\caption{Metabolites that are selected using \textbf{Corrected Lasso} among $\geq$10\% of replications associated with CRC risks and the direction of their marginal association to the CRC risks.}
\label{tab:crc-cl}
\begin{adjustbox}{max width=\textwidth,max totalheight=\textheight,keepaspectratio}
\begin{tabular}{lll}
\hline
  \begin{tabular}[c]{@{}l@{}}Platform \end{tabular} &
 \begin{tabular}[c]{@{}l@{}} Half Min Imputation \end{tabular} &
  \begin{tabular}[c]{@{}l@{}}Multiple Imputation \end{tabular} \\ \hline
  \begin{tabular}[c]{@{}l@{}}NMR \end{tabular}&
  \begin{tabular}[c]{@{}l@{}}N$-$methylnicotinic acid (86\%)(-) \end{tabular} &
  \begin{tabular}[c]{@{}l@{}}Taurine (86\%)(+) \end{tabular} \\ \hline
  \begin{tabular}[c]{@{}l@{}}GCMS \end{tabular} &
  \begin{tabular}[c]{@{}l@{}}2,3$-$Dihydroxybutanoic acid ({\color{black}47}\%)(+)\\
  Phenol, 2,4$-$bis(1,1$-$dimethylethyl)$-$, phosphite (3:1)({\color{black}22\%)}(+)\end{tabular} &
  \begin{tabular}[c]{@{}l@{}}Pseudo uridine penta$-$tms ({\color{black}20}\%)(+)\\2,3$-$Dihydroxybutanoic acid ({\color{black}17}\%)(+)\\Alpha$-$ketoglutarate ({\color{black}15}\%)(+)\\{\color{black} 4,5$-$dihydroxy$-$1,2$-$dithiane (14\%)(+)} \end{tabular} \\ \hline
 
  \begin{tabular}[c]{@{}l@{}}LCMS \\ (AbsQuant)\end{tabular} &
  \begin{tabular}[c]{@{}l@{}}Methionine (46\%)(-)\\ Glucose (29\%)(+)\end{tabular} &
  \begin{tabular}[c]{@{}l@{}}Methionine (59\%)(-)\\ Glucose (59\%)(+)\\iso$-$Leucine (23\%)(+)\\Leucine (15\%)(+)\end{tabular}\\ \hline
 
  \begin{tabular}[c]{@{}l@{}}LCMS \\ (RelQuant)\end{tabular} &
  \begin{tabular}[c]{@{}l@{}} Malate (71\%)(+)\end{tabular} &
  \begin{tabular}[c]{@{}l@{}}Malate (87\%)(+)\end{tabular} \\ \hline
 
  \begin{tabular}[c]{@{}l@{}}Lipidyzer \\ (composition)\end{tabular} &
  \begin{tabular}[c]{@{}l@{}}TAG 44:0(FA16:0) (15\%)(+)\\ TAG 44:0(FA14:0) (14\%)(-)\\ TAG 46:0(FA14:0) (11\%)(-)\\ TAG 46:0(FA16:0) (11\%)(-)\end{tabular} &
  \begin{tabular}[c]{@{}l@{}} \end{tabular} \\ \hline
 
  \begin{tabular}[c]{@{}l@{}}Lipidyzer \\ (concentration)\end{tabular} &
  \begin{tabular}[c]{@{}l@{}}TAG 44:0(FA16:0) (43\%)(-)\\ TAG 44:0(FA14:0) (21\%)(-)\\ TAG 44:1(FA14:0) (11\%)(-)\end{tabular} &
  \begin{tabular}[c]{@{}l@{}} \end{tabular} \\ \hline
\end{tabular}
\end{adjustbox}
\end{table}

\subsection*{Metabolites selected for more than 10\% of times that are associated with both BC and CRC.}


Tables \ref{tab:common-lasso}, \ref{tab:common-lo} and \ref{tab:common-gmus} provide the variables selected from at least 10\% of the replications and the corresponding percentage time of selection for mutual risk factor analysis.
\begin{table}[h]
\centering
\caption{Metabolites that are selected using \textbf{Lasso} among $\geq$10\% of replications associated with both BC and CRC risks and the direction of their marginal association to these two cancer risks.}
\label{tab:common-lasso}
\begin{adjustbox}{max width=\textwidth,max totalheight=\textheight,keepaspectratio}
\begin{tabular}{lll}
\hline
  \begin{tabular}[c]{@{}l@{}}Platform \end{tabular}&
  \begin{tabular}[c]{@{}l@{}}Half Min Imputation \end{tabular}&
 \begin{tabular}[c]{@{}l@{}} Multiple Imputation \end{tabular}\\ \hline
  \begin{tabular}[c]{@{}l@{}}NMR \end{tabular}&
  \begin{tabular}[c]{@{}l@{}}  \end{tabular}&
  \begin{tabular}[c]{@{}l@{}} \end{tabular}\\ \hline
 
  \begin{tabular}[c]{@{}l@{}}GCMS \end{tabular}&
  \begin{tabular}[c]{@{}l@{}}  \end{tabular}&
  \begin{tabular}[c]{@{}l@{}} \end{tabular} \\ \hline
 
  \begin{tabular}[c]{@{}l@{}}LCMS \\ (AbsQuant)\end{tabular} &
  \begin{tabular}[c]{@{}l@{}}Cystine (44\%)(B:-)(C:-)\\ Choline (39\%)(B:+)(C:+)\\ 3HBA (31\%)(B:+)(C:+)\\ Glutamic acid (16\%)(B:+)(C:+)\end{tabular} &
  \begin{tabular}[c]{@{}l@{}}Choline (17\%)(B:+)(C:+)\end{tabular} \\ \hline
 
  \begin{tabular}[c]{@{}l@{}} LCMS \\ (RelQuant)\end{tabular} &
  \begin{tabular}[c]{@{}l@{}} Malate (11\%)(B:-)(C:+) \end{tabular}&
  \begin{tabular}[c]{@{}l@{}}   \end{tabular}\\\hline
 
  \begin{tabular}[c]{@{}l@{}}Lipidyzer \\ (composition)\end{tabular} &
  \begin{tabular}[c]{@{}l@{}}  \end{tabular}&
  \begin{tabular}[c]{@{}l@{}}  \end{tabular}\\\hline
 
  \begin{tabular}[c]{@{}l@{}}Lipidyzer \\ (concentration)\end{tabular} &
  \begin{tabular}[c]{@{}l@{}} \end{tabular} &
  \begin{tabular}[c]{@{}l@{}} \end{tabular} \\ \hline
\end{tabular}
\end{adjustbox}
\end{table}

\begin{table}[h]
\centering
\caption{Metabolites that are selected using \textbf{Lasso Order} among $\geq$10\% of replications associated with both BC and CRC risks and the direction of their marginal association to these two cancer risks.}
\label{tab:common-lo}
\begin{adjustbox}{max width=\textwidth,max totalheight=\textheight,keepaspectratio}
\begin{tabular}{lll}
\hline
  \begin{tabular}[c]{@{}l@{}} Platform \end{tabular}&
  \begin{tabular}[c]{@{}l@{}} Half Min Imputation \end{tabular}&
 \begin{tabular}[c]{@{}l@{}}  Multiple Imputation \end{tabular}\\\hline
 \begin{tabular}[c]{@{}l@{}}  NMR \end{tabular}&
  \begin{tabular}[c]{@{}l@{}} N$-$methylnicotinic acid (48\%)(B:+)(C:-)\end{tabular}&
  \begin{tabular}[c]{@{}l@{}} N$-$methylnicotinic acid (57\%)(B:+)(C:-)\end{tabular} \\ \hline
 
  \begin{tabular}[c]{@{}l@{}} GCMS \end{tabular}&
  \begin{tabular}[c]{@{}l@{}}2,3$-$Dihydroxybutanoic acid ({\color{black}28}\%)(B:-)(C:+) \end{tabular}&
  \begin{tabular}[c]{@{}l@{}}2,3$-$Dihydroxybutanoic acid ({\color{black}17}\%)(B:-)(C:+)\end{tabular} \\\hline
 
  \begin{tabular}[c]{@{}l@{}}LCMS \\ (AbsQuant)\end{tabular} &
  \begin{tabular}[c]{@{}l@{}}Cystine (89\%)(B:-)(C:-)\\ 3HBA (68\%)(B:+)(C:+)\\ Glutamic acid (49\%)(B:+)(C:+)\\ Choline (21\%)(B:+)(C:+)\end{tabular} &
  \begin{tabular}[c]{@{}l@{}}Cystine (99\%)(B:-)(C:-)\\ 3HBA (83\%)(B:+)(C:+)\\ Glutamic acid (78\%)(B:+)(C:+)\\ Pentothenate (28\%)(B:-)(C:+)\\ Urate (17\%)(B:+)(C:+)\\Aspartic Acid (12\%)(B:+)(C:+)\end{tabular} \\\hline
 
  \begin{tabular}[c]{@{}l@{}}LCMS \\ (RelQuant)\end{tabular} &
 \begin{tabular}[c]{@{}l@{}} Malate (30\%) (B:-)(C:+)\end{tabular}&
 \begin{tabular}[c]{@{}l@{}}  \end{tabular} \\ \hline
 
  \begin{tabular}[c]{@{}l@{}}Lipidyzer \\ (composition)\end{tabular} &
  \begin{tabular}[c]{@{}l@{}}TAG 50:5(FA16:1) (15\%)(B:-)(C:-)\\ TAG 44:1(FA12:0) (14\%)(B:-)(C:-)\end{tabular} &
  \begin{tabular}[c]{@{}l@{}}DAG 14:1/18:1 (19\%)(B:+)(C:+)\end{tabular} \\ \hline
 
  \begin{tabular}[c]{@{}l@{}}Lipidyzer \\ (concentration)\end{tabular} &
  \begin{tabular}[c]{@{}l@{}}  \end{tabular}&
  \begin{tabular}[c]{@{}l@{}} \end{tabular}\\ \hline
\end{tabular}
\end{adjustbox}
\end{table}

\begin{table}[h]
\centering
\caption{Metabolites that are selected using \textbf{GMUS} among $\geq$10\% of replications associated with both BC and CRC risks and the direction of their marginal association to these two cancer risks.}
\label{tab:common-gmus}
\begin{adjustbox}{max width=\textwidth,max totalheight=\textheight,keepaspectratio}
\begin{tabular}{lll}
\hline
  \begin{tabular}[c]{@{}l@{}}Platform \end{tabular}&
  \begin{tabular}[c]{@{}l@{}}Half Min Imputation \end{tabular}&
 \begin{tabular}[c]{@{}l@{}} Multiple Imputation \end{tabular}\\ \hline
  \begin{tabular}[c]{@{}l@{}}NMR \end{tabular}&
  \begin{tabular}[c]{@{}l@{}}  \end{tabular}&
  \begin{tabular}[c]{@{}l@{}} \end{tabular} \\\hline
 
  \begin{tabular}[c]{@{}l@{}}GCMS \end{tabular}&
 \begin{tabular}[c]{@{}l@{}}   \end{tabular}&
  \begin{tabular}[c]{@{}l@{}} \end{tabular} \\ \hline
 
  \begin{tabular}[c]{@{}l@{}}LCMS \\ (AbsQuant)\end{tabular} &
  \begin{tabular}[c]{@{}l@{}}Choline (56\%)(B:+)(C:+)\\ Glutamic acid (23\%)(B:+)(C:+)\\ Cystine (12\%)(B:-)(C:-)\end{tabular} &
  \begin{tabular}[c]{@{}l@{}}Choline (63\%)(B:+)(C:+)\\ Glutamic acid (10\%)(B:+)(C:+)\end{tabular} \\ \hline
 
  \begin{tabular}[c]{@{}l@{}}LCMS \\ (RelQuant)\end{tabular} &
  \begin{tabular}[c]{@{}l@{}}  \end{tabular}&
  \begin{tabular}[c]{@{}l@{}}  \end{tabular}\\ \hline
 
  \begin{tabular}[c]{@{}l@{}}Lipidyzer \\ (composition)\end{tabular} &
   \begin{tabular}[c]{@{}l@{}}  \end{tabular}&
  \begin{tabular}[c]{@{}l@{}}  \end{tabular} \\ \hline
 
  \begin{tabular}[c]{@{}l@{}}Lipidyzer \\ (concentration)\end{tabular} &
 \begin{tabular}[c]{@{}l@{}}   \end{tabular}&
  \begin{tabular}[c]{@{}l@{}}  \end{tabular}\\ \hline
\end{tabular}
\end{adjustbox}
\end{table}

\subsection*{Sensitivity analysis using cancer-specific matched controls}

Metabolites that are robustly ($\geq$50\% times selected) associated with BC, CRC, and mutual risks and the direction of their marginal association to the BC/CRC/mutual risks using the corresponding cancer specific matched controls are listed in Tables \ref{tab:my-tablebc}, \ref{tab:my-tablecc} and \ref{tab:my-tablemutual}.

\begin{table}[h]
\centering
\caption{Metabolites that are robustly ($\geq$50\% times selected) associated with BC risks and the direction of their marginal association to the BC risks using BC specific matched controls.}
\label{tab:my-tablebc}
\begin{adjustbox}{max width=\textwidth,max totalheight=\textheight,keepaspectratio}
\begin{tabular}{llll}
\hline
\begin{tabular}[c]{@{}l@{}}Method \end{tabular}& 
\begin{tabular}[c]{@{}l@{}}Platform \end{tabular}&
\begin{tabular}[c]{@{}l@{}}Half Min Imputation\end{tabular} &
\begin{tabular}[c]{@{}l@{}}Multiple Imputation\end{tabular}\\\hline
\begin{tabular}[c]{@{}l@{}}Lasso\end{tabular}&
\begin{tabular}[c]{@{}l@{}}NMR\end{tabular}&
\begin{tabular}[c]{@{}l@{}}N$-$methylnicotinic acid (53\%)(+)\end{tabular} &
\begin{tabular}[c]{@{}l@{}} \end{tabular}\\\hline
\begin{tabular}[c]{@{}l@{}}Lasso \end{tabular}&
\begin{tabular}[c]{@{}l@{}}Lipidyzer \\ (composition)\end{tabular} &
\begin{tabular}[c]{@{}l@{}}TAG 47:0(FA15:0) (54\%)(-)\end{tabular} &
\begin{tabular}[c]{@{}l@{}}DAG 14:1/18:1 (81\%)(+)\end{tabular} \\\hline
\begin{tabular}[c]{@{}l@{}}Lasso \end{tabular}&
\begin{tabular}[c]{@{}l@{}}Lipidyzer \\ (concentration)\end{tabular} &
\begin{tabular}[c]{@{}l@{}}DAG 14:1/18:1 (88\%)(+)\\ TAG 48:0(FA16:0) (59\%)(+)\end{tabular} & 
\begin{tabular}[c]{@{}l@{}} \end{tabular}\\\hline
\begin{tabular}[c]{@{}l@{}}Lasso Order \end{tabular}&
\begin{tabular}[c]{@{}l@{}}GCMS \end{tabular} &
\begin{tabular}[c]{@{}l@{}}Alpha$-$ketoglutarate ({\color{black}100}\%)(-)\end{tabular} &
\begin{tabular}[c]{@{}l@{}} \end{tabular}\\\hline
\begin{tabular}[c]{@{}l@{}} Lasso Order\end{tabular}&
\begin{tabular}[c]{@{}l@{}}NMR\end{tabular} &
\begin{tabular}[c]{@{}l@{}}N$-$methylnicotinic acid (53\%)(+)\end{tabular} &
\begin{tabular}[c]{@{}l@{}}N$-$methylnicotinic acid (65\%)(+)\end{tabular} \\ \hline
\begin{tabular}[c]{@{}l@{}}Lasso Order \end{tabular}&
\begin{tabular}[c]{@{}l@{}}LCMS \\ (AbsQuant)\end{tabular} &
\begin{tabular}[c]{@{}l@{}}Cystine (58\%)(-)\end{tabular} &  
\begin{tabular}[c]{@{}l@{}} \end{tabular} \\\hline
\begin{tabular}[c]{@{}l@{}}Lasso Order \end{tabular} &
\begin{tabular}[c]{@{}l@{}}Lipidyzer \\ (composition)\end{tabular} &
\begin{tabular}[c]{@{}l@{}}TAG 47:0(FA15:0) (62\%)(-) \end{tabular} &
\begin{tabular}[c]{@{}l@{}}DAG 14:1/18:1 (99\%)(+) \end{tabular} \\\hline
\begin{tabular}[c]{@{}l@{}}GMUS \end{tabular}&
\begin{tabular}[c]{@{}l@{}}NMR \end{tabular} &
\begin{tabular}[c]{@{}l@{}}Uracil (95\%)(-)\\Formate (83\%)(+)\end{tabular} &
\begin{tabular}[c]{@{}l@{}}Uracil (97\%)(-)\\Formate (82\%)(+)\end{tabular} \\\hline
\begin{tabular}[c]{@{}l@{}}GMUS \end{tabular}&
\begin{tabular}[c]{@{}l@{}}Lipidyzer \\ (composition)\end{tabular} &
\begin{tabular}[c]{@{}l@{}}TAG 52:2(FA18:2) (60\%)(+)\end{tabular}&
\begin{tabular}[c]{@{}l@{}} \end{tabular}\\\hline
\end{tabular}
\end{adjustbox}
\end{table}
 
\begin{table}[h]
\centering
\caption{Metabolites that are robustly ($\geq$50\% times selected) associated with CRC risks and the direction of their marginal association to the CRC risks using CRC specific matched controls.}
\label{tab:my-tablecc}
\begin{adjustbox}{max width=\textwidth,max totalheight=\textheight,keepaspectratio}
\begin{tabular}{llll}
\hline
\begin{tabular}[c]{@{}l@{}}Method \end{tabular}&
\begin{tabular}[c]{@{}l@{}}Platform \end{tabular}&
\begin{tabular}[c]{@{}l@{}}Half Min Imputation \end{tabular}&
\begin{tabular}[c]{@{}l@{}}Multiple Imputation\end{tabular}\\\hline
\begin{tabular}[c]{@{}l@{}}Lasso \end{tabular}&  
\begin{tabular}[c]{@{}l@{}}LCMS \\ (AbsQuant)\end{tabular}&
\begin{tabular}[c]{@{}l@{}}3HBA (66\%)(+)\\Cystine (60\%)(-)\end{tabular} &
\begin{tabular}[c]{@{}l@{}}  \end{tabular}\\\hline
\begin{tabular}[c]{@{}l@{}}Lasso \end{tabular}&  
\begin{tabular}[c]{@{}l@{}}LCMS \\ (RelQuant)\end{tabular}&
\begin{tabular}[c]{@{}l@{}}Adenosine (52\%)(-)\end{tabular} &
\begin{tabular}[c]{@{}l@{}}Adenosine (90\%)(-) \end{tabular} \\\hline
\begin{tabular}[c]{@{}l@{}}Lasso \end{tabular}&
\begin{tabular}[c]{@{}l@{}}Lipidyzer \\ (composition)\end{tabular} &  
\begin{tabular}[c]{@{}l@{}} \end{tabular}&
\begin{tabular}[c]{@{}l@{}}TAG 48:5(FA18:3) (80\%)(+)\\DAG 14:1/18:1 (78\%)(+) \end{tabular}\\\hline
\begin{tabular}[c]{@{}l@{}}Lasso Order \end{tabular}&
\begin{tabular}[c]{@{}l@{}} NMR \end{tabular}&
\begin{tabular}[c]{@{}l@{}}N$-$methylnicotinic acid (86\%)(-) \end{tabular}&
\begin{tabular}[c]{@{}l@{}}N$-$methylnicotinic acid (94\%)(-) \end{tabular}\\\hline
\begin{tabular}[c]{@{}l@{}}Lasso Order \end{tabular}& 
\begin{tabular}[c]{@{}l@{}}LCMS \\ (AbsQuant)\end{tabular}&
\begin{tabular}[c]{@{}l@{}}3HBA (95\%)(+)\\Cystine (89\%)(-)\end{tabular} &
\begin{tabular}[c]{@{}l@{}}3HBA (98\%)(+)\\Cystine (89\%)(-)\end{tabular} \\\hline
\begin{tabular}[c]{@{}l@{}}Lasso Order \end{tabular}&
\begin{tabular}[c]{@{}l@{}}LCMS \\ (RelQuant)\end{tabular}&  
\begin{tabular}[c]{@{}l@{}} \end{tabular} &
\begin{tabular}[c]{@{}l@{}}Adenosine (90\%)(-)\end{tabular} \\\hline
\begin{tabular}[c]{@{}l@{}}Lasso Order \end{tabular}&  
\begin{tabular}[c]{@{}l@{}}Lipidyzer \\ (composition)\end{tabular} &  
\begin{tabular}[c]{@{}l@{}} \end{tabular}&
\begin{tabular}[c]{@{}l@{}}TAG 48:5(FA18:3) (80\%)(+)\\ DAG 14:1/18:1 (76\%)(+)\end{tabular}  \\\hline
  \begin{tabular}[c]{@{}l@{}} GMUS \end{tabular}&  
  \begin{tabular}[c]{@{}l@{}}LCMS \\ (AbsQuant)\end{tabular}&
  \begin{tabular}[c]{@{}l@{}}Choline (67\%)(+)\end{tabular} & 
  \begin{tabular}[c]{@{}l@{}}\end{tabular}\\\hline
\begin{tabular}[c]{@{}l@{}}GMUS \end{tabular}&  
\begin{tabular}[c]{@{}l@{}}LCMS \\ (RelQuant)\end{tabular}& 
\begin{tabular}[c]{@{}l@{}} \end{tabular}&
\begin{tabular}[c]{@{}l@{}}Adenosine (90\%)(-)\end{tabular} \\\hline
\begin{tabular}[c]{@{}l@{}}GMUS \end{tabular}&  
\begin{tabular}[c]{@{}l@{}}Lipidyzer \\ (composition)\end{tabular} &
\begin{tabular}[c]{@{}l@{}} \end{tabular}&
\begin{tabular}[c]{@{}l@{}}PC 16:0/18:2 (55\%)(-)\end{tabular}  \\\hline
\end{tabular}
\end{adjustbox}
\end{table}

\begin{table}[h]
\centering
\caption{Metabolites that are robustly ($\geq$50\% times selected) associated with both BC and CRC risks and the direction of their marginal association to these two cancer risks using specific cancer controls.}
\label{tab:my-tablemutual}
\begin{adjustbox}{max width=\textwidth,max totalheight=\textheight,keepaspectratio}
\begin{tabular}{llll}
\hline
\begin{tabular}[c]{@{}l@{}}Method \end{tabular}& 
\begin{tabular}[c]{@{}l@{}}Platform \end{tabular}&
\begin{tabular}[c]{@{}l@{}}Half Min Imputation\end{tabular} &
\begin{tabular}[c]{@{}l@{}}Multiple Imputation\end{tabular}\\\hline

 \begin{tabular}[c]{@{}l@{}}Lasso \end{tabular} &
 \begin{tabular}[c]{@{}l@{}}Lipidyzer \\ (composition)\end{tabular} & 
 \begin{tabular}[c]{@{}l@{}}  \end{tabular} &
 \begin{tabular}[c]{@{}l@{}} DAG 14:1/18:1 (58\%)(B:+)(C:+) \end{tabular}\\\hline
 \begin{tabular}[c]{@{}l@{}}Lasso Order \end{tabular}&
 \begin{tabular}[c]{@{}l@{}}  NMR \end{tabular}& 
 \begin{tabular}[c]{@{}l@{}} \end{tabular}&
\begin{tabular}[c]{@{}l@{}} N$-$methylnicotinic acid (56\%)(B:+)(C:-)\end{tabular} \\\hline
\begin{tabular}[c]{@{}l@{}}Lasso Order \end{tabular} &
\begin{tabular}[c]{@{}l@{}}LCMS \\ (AbsQuant)\end{tabular} &
\begin{tabular}[c]{@{}l@{}}Cystine (89\%)(B:-)(C:-)\\ 3HBA (68\%)(B:+)(C:+)\end{tabular} &
\begin{tabular}[c]{@{}l@{}}Cystine (52\%)(B:-)(C:-)\end{tabular}\\\hline
\begin{tabular}[c]{@{}l@{}}GMUS \end{tabular}&
\begin{tabular}[c]{@{}l@{}}LCMS \\ (AbsQuant)\end{tabular} &
\begin{tabular}[c]{@{}l@{}}Choline (56\%)(B:+)(C:+)\end{tabular} &
\begin{tabular}[c]{@{}l@{}} \end{tabular}\\\hline
\end{tabular}
\end{adjustbox}
\end{table}

\end{document}